\documentclass[twocol]{ametsoc}
\pdfoutput=1
\usepackage{widetext} 

\journal{jas}

\bibpunct{(}{)}{;}{a}{}{,}

\title{Vertically Sheared Horizontal Flow-Forming Instability in Stratified Turbulence: \\ Analytical Linear Stability Analysis of Statistical State Dynamics Equilibria}
\authors{Joseph G. Fitzgerald\correspondingauthor{Joseph G. Fitzgerald, Department of Earth and Planetary Sciences, Harvard University, Geological Museum 404, 24 Oxford Street, Cambridge, MA 02138} and Brian F. Farrell}

\affiliation{Department of Earth and Planetary Sciences, Harvard University, Cambridge, Massachusetts} 

\email{jfitzgerald@fas.harvard.edu}

\abstract{Vertically banded zonal jets are frequently observed in weakly or non-rotating stratified turbulence, with
the quasi-biennial oscillation in the equatorial stratosphere and the ocean's equatorial deep jets being two examples.
Explaining the formation of jets in stratified turbulence is a fundamental problem in geophysical fluid dynamics. 
Statistical state dynamics (SSD) provides powerful methods for analyzing turbulent systems exhibiting emergent organization, such as banded jets. 
In SSD, dynamical equations are written directly for the evolution of the turbulence statistics, enabling direct analysis of the statistical interactions between the incoherent component of the turbulence and the coherent large-scale structure component that underlie jet formation. 
A second-order closure of SSD, known as S3T, has previously been applied to show that meridionally banded jets emerge in barotropic $\beta$-plane turbulence via a statistical instability referred to as the zonostrophic instability.
Two-dimensional Boussinesq turbulence provides a simple model of non-rotating stratified turbulence analogous to the $\beta$-plane model of planetary turbulence. 
Jets known as vertically sheared horizontal flows (VSHFs) often emerge in simulations of Boussinesq turbulence, but their dynamics is not yet clearly understood. 
In this work S3T analysis of the zonostrophic instability is extended to study VSHF emergence in two-dimensional Boussinesq turbulence using an analytical formulation of S3T amenable to perturbation stability analysis. 
VSHFs are shown to form via an instability that is analogous in stratified turbulence to the zonostrophic instability in $\beta$-plane turbulence. 
This instability is shown to be strikingly similar to the zonostrophic instability, suggesting that jet emergence in both geostrophic and non-rotating stratified turbulence may be understood as instances of the same generic phenomenon.}

\begin{document}

\maketitle

\section{Introduction}
Coherent zonal jets are a common feature of geostrophic turbulence. 
The meridionally banded zonal winds of Jupiter \citep{Vasavada:2005gs} and the striations of the Earth's midlatitude oceans \citep{Maximenko:2005jf} provide striking examples. 
Zonal jets also emerge in laboratory experiments and numerical simulations modeling the planetary turbulence regime \citep{WILLIAMS:1978va,Huang:1998tw,Read:2007bg,Galperin:2017tg}. 
The barotropic $\beta$-plane system serves as a paradigmatic model for zonal jet emergence in planetary turbulence due to its simplicity as well as its role in the problem's history \citep{Rhines:1975uu}.  
In barotropic and related models of the Jovian jets the jet formation and maintenance mechanisms are often presumed to occur in the planet's shallow stably stratified weather layer, while the jet dynamics permit deep structure (first or external mode), with a finite characteristic depth being presumed in the case of equivalent barotropic models. 
Mechanisms hypothesized to originate in the planetary interior, such as the production of differential rotation by deep columnar convection \citep{Busse:1976ux}, have also been advanced to explain Jupiter's jets. 
Recent satellite observations have revealed that the banded winds have deep structure \citep{Kaspi:2018hm}.
However, given that deep structure may arise from shallow dynamics \citep{Showman:2006jr,Farrell:2017ed} the mechanistic origin of the Jovian jets remains uncertain. 

Organization of geostrophic turbulence into zonal jets is sometimes referred to as `zonation' \citep{Galperin:2006ve} and the mechanism giving rise to zonation is sometimes referred to as the `zonostrophic instability' \citep{Srinivasan:2012im}. 
The zonostrophic instability is a statistical instability in which weak jets arising randomly from turbulent fluctuations or initial conditions break the statistical homogeneity of geostrophic turbulence resulting in the organization of the turbulent Reynolds stresses in a manner such that these stresses drive the jets. 
The instability is intrinsically statistical because the turbulent background upon which it occurs is stochastically fluctuating. 
The interactions between the jets and the turbulence fluctuate in time and space so that mutual reinforcement of the jets and the Reynolds stresses occurs in statistical average but not at each instant or location. 
Because it is intrinsically statistical, analytical solution for the mechanisms and structures giving rise to the zonostrophic instability is not possible using individual realizations, essentially due to the presence of turbulent fluctuations in the realizations. 
When dynamics are instead formulated for the statistical state of the turbulence, an approach referred to as statistical state dynamics (SSD), the obscuring impediment of turbulent fluctuations is eliminated and the zonostrophic instability assumes the form of a canonical linear instability amenable to the familiar analytical techniques of dynamical systems analysis \citep{FARRELL:2014we}. 

Organization of turbulence into persistent zonal jets also occurs in weakly- and non-rotating stratified turbulence. Vertically banded (or `stacked') jets known as equatorial deep jets are observed in all equatorial ocean basins below approximately 1000 m depth and consist of alternating eastward and westward zonal jets with a spacing of approximately 500 m \citep{Youngs:2015gh}. 
The quasi-biennial oscillation of the equatorial stratosphere provides another example in which the vertically banded structure takes the form of regularly descending easterly and westerly jets \citep{Baldwin:2001wi}. 
Laboratory models of non-rotating stratified turbulence in a reentrant annulus also develop banded jets similar to the quasi-biennial oscillation \citep{Plumb:1978cj}.  

The analog of the $\beta$-plane system appropriate for modeling stacked jet formation in stratified turbulence is the stably stratified Boussinesq system. 
Like the $\beta$-plane system, the Boussinesq system does not generate turbulence spontaneously in the absence of an externally forced jet, so turbulence in these systems is traditionally maintained by a stochastic parameterization accounting for exogenous forcing of the turbulence.
Numerical simulations of Boussinesq turbulence frequently develop strong vertically banded horizontal jets \citep{Laval:2003ko,Waite:2004bt,WAITE:2005bj,Brethouwer:2007uk,Marino:2014id,Rorai:2015bw,Herbert:2016cc,Kumar:2017ie}. These jets, often referred to as vertically sheared horizontal flows (VSHFs) or shear modes, develop in both 2D and 3D turbulence and in both non-rotating and weakly-rotating regimes \citep{Smith:2001uo,Smith:2002wg}.  
In previous work we showed that, in 2D stratified turbulence, VSHFs form via a statistical instability of homogeneous stratified turbulence analogous to the zonostrophic instability \citep{Fitzgerald:2016ux}.
We refer to this instability, which belongs to a larger class of SSD instabilities that includes the zonostrophic instability, as the VSHF-forming instability. 

Because the underlying instability is due to statistical organization of the turbulence, the zonostrophic and VSHF-forming instabilities have analytical expression in the SSD of turbulence, rather than in the dynamics of individual turbulent realizations. 
SSD refers to any theoretical approach to the analysis of fluctuating chaotic systems in which equations of motion are formulated directly for statistical variables of the system rather than for the detailed system state. 
For example, the Fokker-Planck equation is an SSD written for the time evolution of the probability density function of the state of any system whose realizations evolve according to a stochastic differential equation. 
The Fokker-Planck equation is an exact SSD, so that the statistical predictions of the Fokker-Planck equation correspond exactly to the statistics obtained by averaging over realizations of the stochastic differential equation. 
However, for systems of practical interest the Fokker-Planck equation cannot be solved numerically due to the extremely high dimension of its state space. 
Stochastic structural stability theory (S3T) \citep{Farrell:2003ud} provides an approximate SSD, closed at second order, that is amenable to numerical solution and theoretical analysis and therefore provides an attractive system for studying the zonostrophic instability and the VSHF-forming instability. 

Recent progress in the application of SSD has resulted from the realization that second-order closure of the SSD comprises the fundamental mechanisms underlying the dynamics of anisotropic turbulence dominated by large coherent structures. 
To obtain the second-order S3T closure, the dynamical variables of the flow are decomposed into two components: a coherent component and an incoherent component. 
For example, in the present work we take the coherent component to be the horizontal mean state and the incoherent component to be the perturbations relative to this mean. 
In the equations of motion of the coherent component all nonlinear interactions are kept intact. 
In the equations of motion of the incoherent component the nonlinear interactions between the coherent and incoherent components are retained, but the self-interactions of the incoherent component are not retained consistent with S3T constituting a canonical second-order closure \citep{Herring:1963gx}. 
The dynamics of the incoherent component is then equivalent to linear evolution about the instantaneous coherent flow. 
The incoherent component feeds back on the coherent component via the Reynolds stresses and buoyancy fluxes. 
S3T is appropriate for analyzing turbulent systems in which the fundamental underlying mechanism is spectrally nonlocal interaction between coherent large-scale structure and incoherent smaller scale turbulence. 
Absence of perturbation-perturbation nonlinearity in the dynamics of the incoherent component of the turbulence in S3T dynamics precludes mechanisms based on a spectrally local turbulent cascade. 

The state variables of S3T are the mean state of the turbulence (the first cumulant, which is the coherent component) and the covariance of the perturbations from the mean state (the second cumulant, which is the incoherent component). 
The mean and the covariance interact quasilinearly (QL) within the second-order closure due to the absence of the self-interactions of the perturbations. 
S3T, and the related second-order closure referred to as CE2 (for second-order cumulant expansion) \citep{Marston:2008gx}, has been successfully applied to study many different turbulent systems that exhibit large-scale coherent structure. 
Even though nonlinearity is highly restricted in QL dynamics, the results of QL and S3T simulations have demonstrated that QL dynamics correctly reproduces the inhomogeneous structure observed in simulations made using barotropic, shallow-water, and two-layer models of planetary turbulence \citep{Farrell:2003ud,Farrell:2007fq,Farrell:2008fd,Farrell:2009cq,Farrell:2009iu,Marston:2010ew,Marston:2012co,Srinivasan:2012im,Tobias:2013hk,Bakas:2013bo,Constantinou:2014fh,Bakas:2014gf,Constantinou:2016fp,Farrell:2017ed}.
These results imply that QL dynamics comprises the physical mechanisms responsible for the formation and maintenance of the equilibrium statistical state of anisotropic turbulence dominated by incoherent turbulence interacting with large-scale coherent structures. 
S3T has also been applied to analyze the interaction of turbulence with large-scale coherent structure in the drift wave-zonal flow plasma system \citep{Farrell:2009dt,Parker:2013hy}, unstratified 2D flow \citep{Bakas:2011bt}, rotating magnetohydrodynamics \citep{Tobias:2011cn,Squire:2015kb}, and the turbulence of stable ion-temperature-gradient modes in plasmas \citep{StOnge:2017tu}. 

Zonal jet emergence in barotropic $\beta$-plane turbulence has been analyzed in depth using S3T. 
Early applications of S3T \citep{Farrell:2003ud,Farrell:2007fq} showed that, for a broad range of parameter values, zonal jets form via the instability referred to as the zonostrophic instability.  
The primary mechanism of jet growth was shown to be spectrally nonlocal transfer of energy from the perturbations into the jets, with the spectrally local incoherent cascade being inessential for the observed jet formation. 
The analytical framework of S3T has since been extended to enable analysis of the jet formation instability in unbounded turbulence using a differential representation \citep{Srinivasan:2012im} as well as the emergence of non-zonal coherent structures \citep{Bernstein:2010gr,Bakas:2013bo} and their coexistence with coherent zonal jets \citep{Constantinou:2016fp}. 
The predictions of S3T and CE2 have been verified through comparison with fully nonlinear simulations \citep{Tobias:2013hk,Bakas:2014gf,Constantinou:2014fh}. 
S3T has also been used to demonstrate that zonal jets can be analyzed within the mathematical and conceptual framework of pattern formation \citep{Parker:2014ui,Bakas:2017uh}. 
Of particular relevance to the present study, S3T has been applied to analyze the mechanism of the zonostrophic instability in great detail, including determining the contribution of specific physical processes, such as shear straining and Rossby wave propagation, to the wave-mean flow interaction that underlies the zonostrophic instability \citep{Bakas:2013ft,Bakas:2015iy}.

Wave-mean flow interactions similar to those which underlie the zonostrophic instability have also been proposed as the drivers of vertically banded jets in stratified turbulence. 
Wave-mean flow interactions between the zonal flow and gravity waves propagating upward from the troposphere underpin the conventional mechanistic explanation of the quasi-biennial oscillation \citep{Holton:1972vm,Plumb:1977fy}. 
In the case of the equatorial deep jets, a number of theoretical explanations have been suggested for their existence, including direct driving by surface winds \citep{Wunsch:1977jl,McCrearyJr:1984bb}, an instability of finite-amplitude equatorial waves \citep{Hua:2008wd}, and nonlinear cascade of baroclinic mode energy in the equatorial region \citep{Salmon:1982up}.
However, recent realistic numerical simulations \citep{Ascani:2015dd} corroborate earlier theoretical analysis \citep{Muench:1999dy} arguing that the jets instead result from wave-mean flow interaction. 
Despite the ubiquity of VSHFs in simulations of stratified turbulence, fewer mechanisms have been proposed for their existence. 
A commonly advanced idea is that resonant and near-resonant interactions among gravity waves may play an important role \citep{Smith:2001uo,Smith:2002wg}.
Recently, we have applied S3T in its finite difference matrix formulation to show that in 2D stratified turbulence the VSHF emerges as a result of an S3T instability analogous to the zonostrophic instability, and to analyze how the VSHF is equilibrated and maintained at finite amplitude \citep{Fitzgerald:2016ux}.

Here we carry out an S3T analysis of VSHF emergence in 2D stratified turbulence that complements our previous work by taking advantage of the differential linearized approach to analyzing S3T instabilities first developed by \citet{Srinivasan:2012im} in the context of the zonostrophic instability. 
Our previous work primarily addressed the structure and maintenance mechanism of finite amplitude VSHFs \citep{Fitzgerald:2016ux}, and used the traditional matrix implementation of S3T appropriate for this purpose \citep{Farrell:2003ud}. 
This analysis expands on that of our previous work in several ways. 
First, use of the differential approach enables characterization of the VSHF-forming instability in terms of a closed-form dispersion relation for the instability growth rate in which the dependence on parameters such as the stratification strength is explicit and which is amenable to asymptotic analysis. 
This approach also enables the application of techniques developed by \citet{Bakas:2013ft} and \citet{Bakas:2015iy}, in the context of the zonostrophic instability, to analyze the wave-mean flow feedback mechanism of the VSHF-forming instability in detail. 
Here we apply these analytical tools to study the VSHF-forming instability mechanism and its relation to the structure of the underlying turbulence, and to determine the roles of various physical processes, such as gravity wave dynamics and shear straining of the vorticity field, in the instability mechanism. 
S3T, and the differential linear approach in particular, allows these determinations to be made straightforwardly and with greater clarity than would be possible through interpretation of nonlinear simulations. 

The rest of the paper is structured as follows. 
In Section \ref{sec:NLphenom} we introduce the fully nonlinear equations of motion (NL) for the 2D stochastically maintained Boussinesq system and its QL counterpart and show the results of example simulations illustrating the phenomenon of VSHF emergence and the degree to which the QL and S3T systems accurately capture the VSHF behavior. 
In Section \ref{sec:S3Tderivation} we formulate the S3T equations. 
In Section \ref{sec:linearization} we apply the differential linearized S3T approach to analyze the linear stability of homogeneous stratified turbulence and derive a dispersion relation for the growth rate of the VSHF-forming instability. 
We also derive a dispersion relation for a related S3T instability governing the emergence of horizontal mean buoyancy layers which we refer to as the buoyancy layering instability. 
In Section \ref{sec:dispersion} we apply these dispersion relations to analyze how the VSHF-forming and buoyancy layering instabilities depend on the parameters and on the structure of the underlying turbulence.
In Section \ref{sec:stabilityboundary} we analyze the stability boundary, or neutral curve, of the VSHF-forming instability and compare the predictions of S3T to the results of NL simulations. 
In Sections \ref{sec:feedback} and \ref{sec:processes} we analyze the wave-mean flow feedback mechanisms of the VSHF-forming and buoyancy layering instabilities in detail. 
We provide a summary and discussion in Section \ref{sec:discussion}.

\section{Emergence of Horizontal Mean Structure in 2D Stratified Turbulence} \label{sec:NLphenom}

\subsection{NL System}

We study VSHF formation in 2D stably stratified Boussinesq turbulence maintained by homogeneous stochastic excitation. 
For our theoretical analysis we use a domain that is unbounded in both directions, and for our numerical simulations we use a doubly periodic domain of unit aspect ratio. 
The equations of motion of the NL system are
\begin{align}
\partial_t \zeta &= -J(\psi,\zeta)+\partial_x b +\sqrt{\varepsilon} \xi^{\zeta}-r\zeta'-r_m\overline{\zeta}+\nu \Delta \zeta,  \label{eq:NLpredecomp1} \\
\partial_t b &= -J(\psi,b)-(\partial_x \psi)N_0^2 + \sqrt{\varepsilon} \xi^b -rb'-r_m\overline{b}+\nu\Delta b, \label{eq:NLpredecomp2} 
\end{align}
in which $x$ and $z$ are the horizontal and vertical coordinates, $\Delta=\partial^2_{xx}+\partial^2_{zz}$ is the Laplacian, $J(f,g)=(\partial_x f)(\partial_z g) - (\partial_z f)(\partial_x g)$ is the Jacobian, $\psi$ is the streamfunction satisfying $(-\partial_z \psi,\partial_x\psi)=(u,w)$ where $u$ and $w$ are the horizontal and vertical velocity components, $\zeta$ is the vorticity defined as $\zeta=\partial_x w - \partial_z u=\Delta \psi$ and $b$ is the buoyancy. 
We denote the horizontal mean operator by an overbar and perturbations from the mean by a prime. 
The stochastic excitations of the vorticity and buoyancy fields are denoted by $\xi^{\zeta}$ and $\xi^b$. 

The parameters of the system are the strength of the stochastic excitation, $\varepsilon$, the constant background buoyancy frequency, $N_0^2$, the Rayleigh drag coefficients for the perturbations, $r$, and for the mean fields, $r_m$, and the viscosity, $\nu$. 
Although fundamental studies of stratified turbulence typically do not include large-scale dissipation such as Rayleigh drag, we do so here both to connect our analysis more closely to the well-studied $\beta$-plane turbulence system, as well as to model the effects of turbulent dissipation by processes that are unresolved in our 2D system. 
Turbulent dissipation is conventionally parameterized as diffusive and therefore damps the large scales less strongly. 
As a simplified model of such scale-dependent dissipation we use different Rayleigh drag coefficients for the mean and perturbation fields, with the mean coefficient being the smaller of the two.
We set the values of the dissipation parameters $r$, $r_m$, and $\nu$, so that the buoyancy and velocity/vorticity fields are damped with equal strength, following standard practice in previous studies of VSHFs. 
Viscosity is chosen to be small and is included to ensure numerical convergence. 

Anticipating the formation of horizontal mean structure we write (\ref{eq:NLpredecomp1})-(\ref{eq:NLpredecomp2}) in Reynolds-decomposed form in which the averaging operator is the horizontal mean.  
For convenience we denote the horizontal mean velocity and buoyancy by capital letters so that $\overline{u}\equiv U$ and $\overline{b}\equiv B$. 
The Reynolds decomposed equations are
\begin{multline}
\partial_t \zeta' = - U\partial_x \zeta' +\partial^2_{zz}U\partial_x \psi' +\partial_x b'+\sqrt{\varepsilon} \xi^{\zeta} \\ 
-r\zeta' + \nu \Delta \zeta' + \text{EENL}^{\zeta}, \label{eq:NL1} 
\end{multline}
\begin{multline}
\partial_t b' = - U\partial_x b' -(N_0^2+\partial_z B)\partial_x \psi' + \sqrt{\varepsilon} \xi^b \\ 
-rb' + \nu \Delta b' +\text{EENL}^b, \label{eq:NL2} 
\end{multline}
\begin{align}
\partial_ t U &= - \partial_z \overline{u'w'} -r_m U +  \nu \partial^2_{zz} U,  \label{eq:NL3} \\ 
\partial_t B &= - \partial_z \overline{w'b'}-r_m B  + \nu \partial^2_{zz} B,  \label{eq:NL4}
\end{align}
where EENL$^{\zeta}$ and EENL$^b$ denote the eddy-eddy nonlinear terms in the perturbation vorticity and buoyancy equations which are produced by the advection of perturbations by perturbations and which are given by the expressions
\begin{align}
\text{EENL}^{\zeta} & \equiv - \left[ J(\psi',\Delta \psi')-\overline{J(\psi',\Delta\psi')}\right], && \label{eq:NL5}\\
\text{EENL}^{b} & \equiv - \left[ J(\psi',b')-\overline{J(\psi',b')}\right]. && \label{eq:NL6}
\end{align}
Stochastic excitation appears in the perturbation equations (\ref{eq:NL1})-(\ref{eq:NL2}) and not in the mean equations (\ref{eq:NL3})-(\ref{eq:NL4}) because we choose the excitation to have zero horizontal mean. 

\begin{figure}[t]
\centerline{\includegraphics[scale=1]{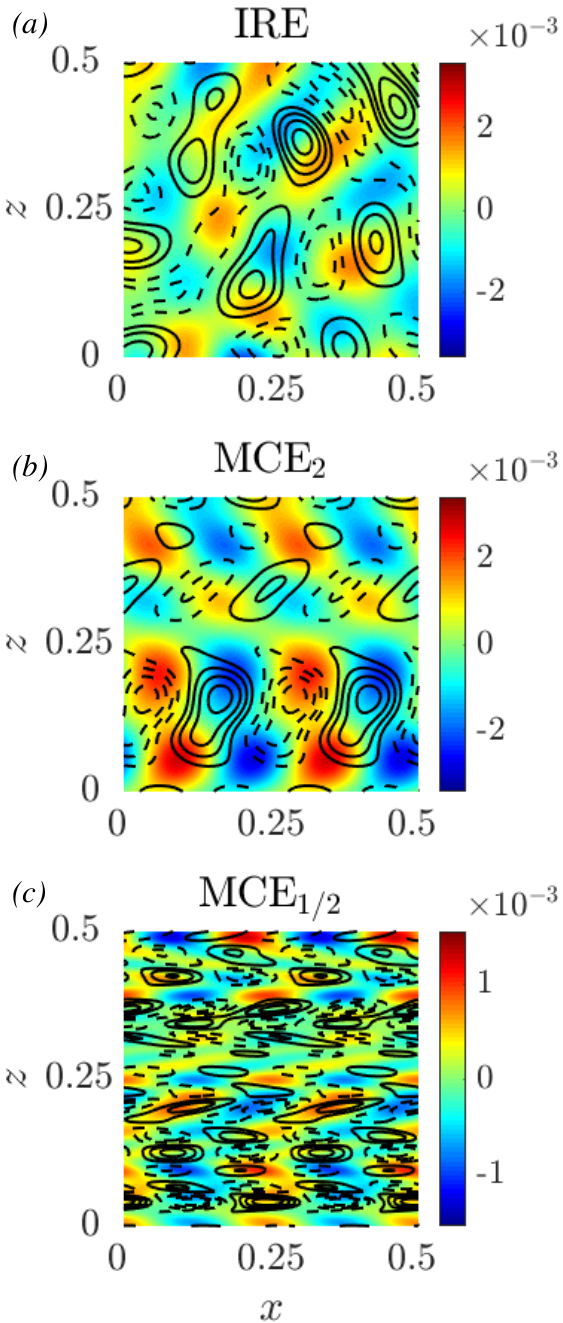}}
\caption{Physical space realizations of the excitation structures IRE (a) and MCE (b,c). The domain is square and doubly periodic with linear dimension $L=1$, and a quarter of the domain is shown. In this example, IRE excites perturbations in a narrow ring of wavenumbers centered at $k_e/(2\pi L)=4$ and MCE excites the horizontal wavenumber component $k_0/(2\pi L)=4$ with a Gaussian covariance structure in the vertical direction. The autocorrelation lengthscales for the MCE buoyancy excitation are $k_0\ell_c = 2$ (b) and $k_0\ell_c = 0.5$ (c). Colors (contours) show the buoyancy (vorticity) excitation. The excitation is shown in normalized form so that $\text{max}(\xi^{\zeta})=1$ but the relative amplitudes of $\xi^{\zeta}$ and $\xi^b$ are preserved. The relative amplitudes of $\xi^{\zeta}$ and $\xi^b$ are set to satisfy the condition of equal kinetic and potential energy injection for $N_0^2=100$ (see Section 4). Solid (dashed) contours indicate positive (negative) vorticity excitation, and the contour levels are $\pm \left\{0.2,0.4,0.6,0.8 \right\}$.} \label{fig:excitation_phys} 
\end{figure}

To complete the formulation of the NL system it remains to specify the stochastic excitation. 
We analyze the conventionally studied case of turbulence maintained by statistically stationary excitation which is white in time and has a prescribed covariance structure in space. 
The two-point, two-time covariance function of the vorticity excitation, $\xi^{\zeta}$, is thus given by 
\begin{equation}
\langle \xi^{\zeta}_1(t_i) \xi_2^{\zeta}(t_j) \rangle \equiv \delta(t_i-t_j) \Xi(\vec{x}_1,\vec{x}_2),
\end{equation}
where we have introduced the notation $\xi^{\zeta}_{1,2}(t) = \xi^{\zeta}(\vec{x}_{1,2},t)$. Similarly, we have 
\begin{align}
\langle \xi^{b}_1(t_i) \xi_2^{b}(t_j)  \rangle & \equiv \delta(t_i-t_j) \Theta (\vec{x}_1,\vec{x}_2), \\
\langle \xi^{\zeta}_1(t_i) \xi_2^{b}(t_j) \rangle & \equiv \delta(t_i-t_j) G^{\zeta} (\vec{x}_1,\vec{x}_2), \\
\langle \xi^{b}_1(t_i) \xi_2^{\zeta}(t_j) \rangle & \equiv \delta(t_i-t_j) G^b (\vec{x}_1,\vec{x}_2),
\end{align}
for the covariance of $\xi^b$ and the covariances between $\xi^{\zeta}$ and $\xi^b$. 
The structure of the excitation in space is determined by the choice of the functions $\Xi$, $\Theta$, $G^{\zeta}$, and $G^b$ (see Appendix B for a concrete example). 
In this work we use two excitation distributions: Isotropic ring excitation (IRE) in which the excitation injects energy into a narrow ring in wavenumber space, and monochromatic excitation (MCE) in which the excitation injects energy into a single horizontal wavenumber component with a Gaussian covariance structure in the vertical direction. 
Mathematical descriptions of IRE and MCE are provided in Section \ref{sec:dispersion} and examples of snapshot realizations in physical space are shown in Figure \ref{fig:excitation_phys}. 
In all cases we choose the excitation to be statistically homogeneous so that $\Xi$, $\Theta$, $G^{\zeta}$ and $G^b$ depend on $\vec{x}_1$ and $\vec{x}_2$ only in the combination $\vec{x}_1-\vec{x}_2$. 
We also note that, because the point labels $\vec{x}_1$ and $\vec{x}_2$ are arbitrary, the covariance functions obey the symmetry relations
\begin{align}
\Xi(\vec{x}_1-\vec{x}_2) &= \Xi(\vec{x}_2-\vec{x}_1), \\
\Theta(\vec{x}_1-\vec{x}_2) &= \Theta(\vec{x}_2-\vec{x}_1), \\
G^{\zeta}(\vec{x}_1-\vec{x}_2) &= G^b(\vec{x}_2-\vec{x}_1). \label{eq:GzetaGb_exchange}
\end{align}


\begin{figure*}[t]
\centerline{\includegraphics[scale=1]{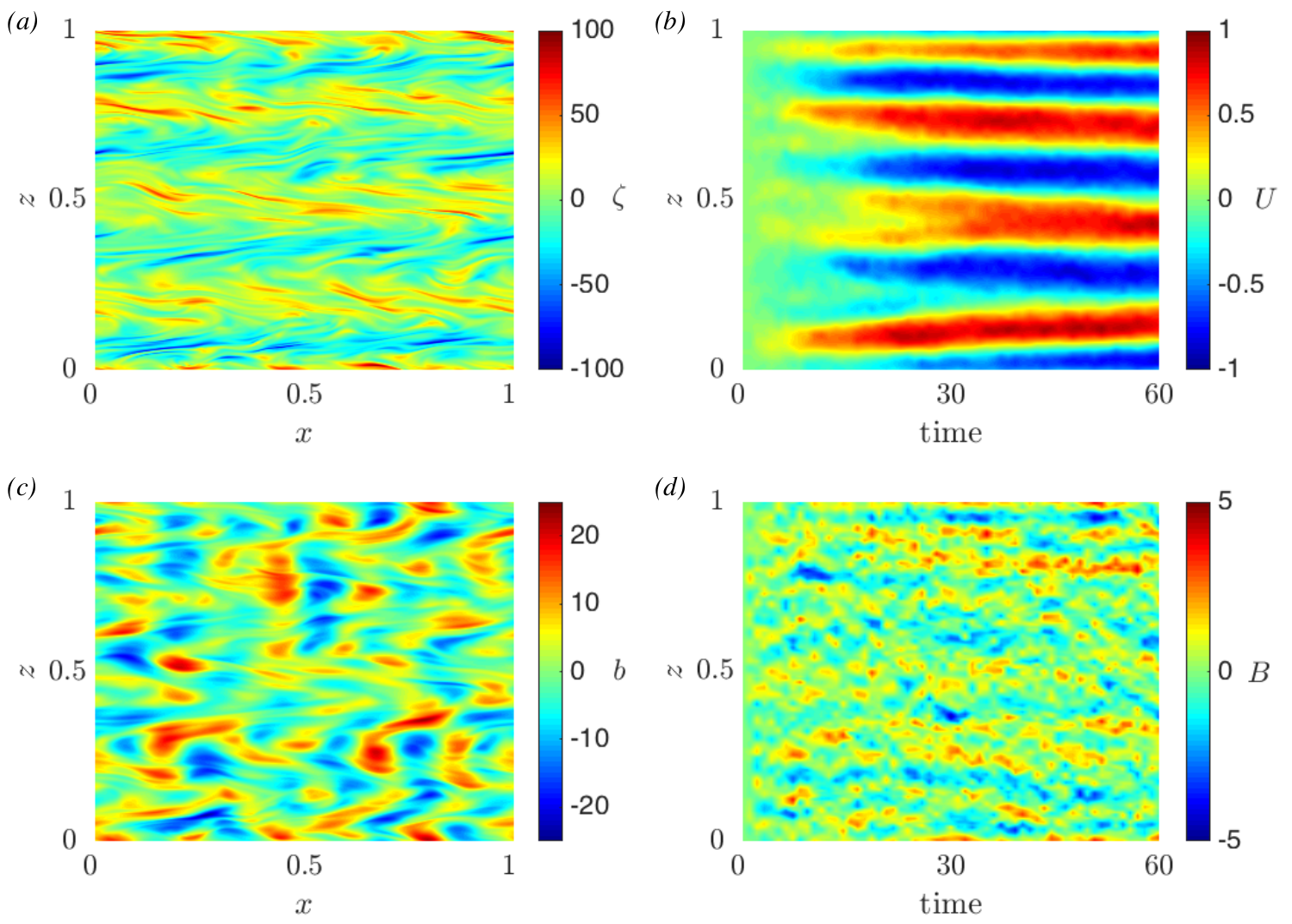}}
\caption{Emergence of a VSHF in the NL system in which turbulence is maintained with IRE. Panels show (a) the final state of the vorticity field, (b) the time evolution of the horizontal mean velocity, (c) the final state of the buoyancy field, and (d) the time evolution of the horizontal mean buoyancy. 
The NL system spontaneously develops vertical banding in the vorticity field associated with the development of a strong VSHF. 
The buoyancy is also organized into more unsteady layered structures that are not apparent in snapshots but are revealed by horizontal averaging. 
For comparison with the results of Section \ref{sec:dispersion}, the nondimensional parameters used are $\varepsilon=177$, $N_0^2=10^3$, $r_m=0.1$, and $\nu=0.03$, and the nondimensional wavenumber of the emergent VSHF is $m=1/\sqrt{2}$. 
Dimensional parameters and simulation details can be found in the text.} \label{fig:NLphenom}
\end{figure*}

Figure \ref{fig:NLphenom} shows an example simulation of the NL system in which a VSHF forms. 
Equations (\ref{eq:NL1})-(\ref{eq:NL4}) were solved in a doubly periodic domain of unit aspect ratio using the finite-difference fluid solver DIABLO \citep{Taylor:2008um} with a resolution of 512 grid points in the $x$ and $z$ directions. 
In dimensional units the domain size is $L=1$ m and the parameters used are $r=1$ s$^{-1}$, $r_m=0.1$ s$^{-1}$, $N_0^2=10^3$ s$^{-2}$, $\nu=2.4\times10^{-5}$ m$^2$ s$^{-1}$, $\varepsilon=0.14$ m$^2$ s$^{-3}$.
Turbulence is maintained with IRE centered on the ring of wavenumbers with $k_e/(2\pi) = 4\sqrt{2}$ m$^{-1}$. 
Figure \ref{fig:NLphenom} (a) shows a snapshot of the vorticity field after 60 s of spin up, illustrating the emergent vertical banding. 
The bands coincide with the shear regions of an energetic VSHF, $U$, the time evolution of which is shown in Figure  \ref{fig:NLphenom} (b). 
The instantaneous buoyancy field, shown in Figure \ref{fig:NLphenom} (c), does not show obvious vertical banding. 
However, time evolution of the horizontal mean buoyancy, $B$, reveals mean layered structures that are too weak to be visible in the instantaneous snapshots but are persistent over several mean damping times (Figure \ref{fig:NLphenom} (d)). 
We refer to such layered buoyancy structures as horizontal mean buoyancy layers. 

\begin{figure}[t]
\centerline{\includegraphics[scale=1.]{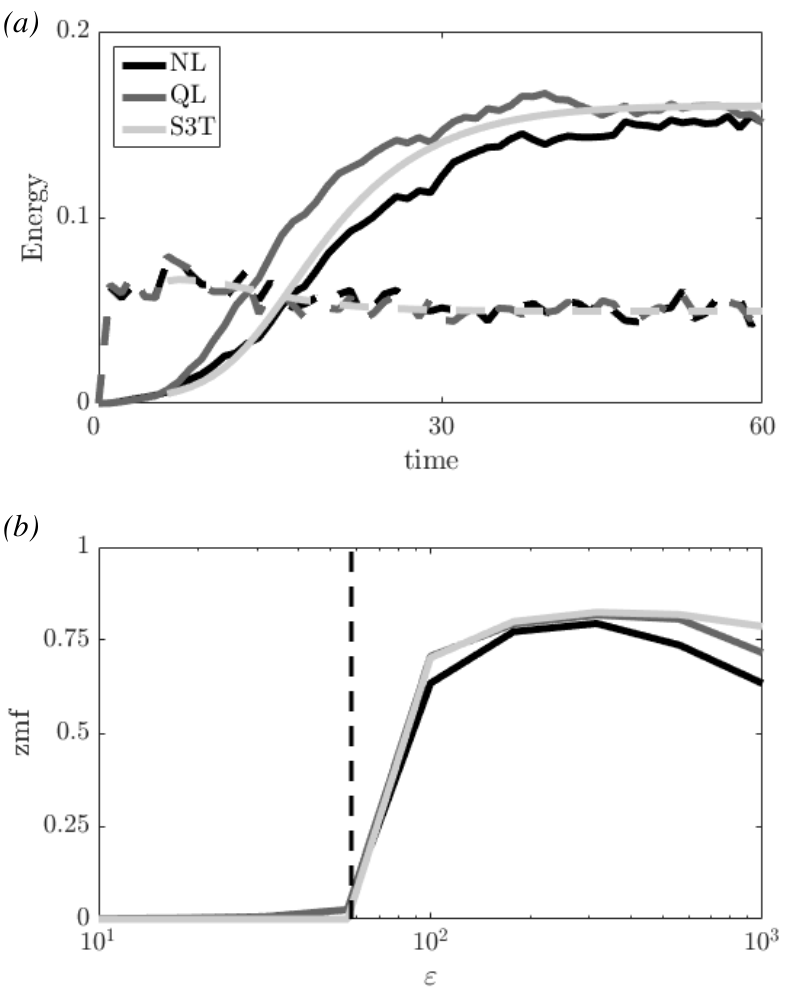}}
\caption{Comparison of VSHF emergence diagnostics in simulations of the NL, QL, and S3T systems. Parameters and numerical details are as in Figure \ref{fig:NLphenom} unless otherwise specified. Panel (a) shows the dimensional total energy of the perturbation field (dashed) and the dimensional kinetic energy of the VSHF (solid) as functions of time for nondimensional excitation strength $\varepsilon=177$. Panel (b) shows the fraction of the total energy contained in the VSHF after a spin-up period as a function of $\varepsilon$. The S3T system captures the behavior of the NL and QL systems. In each system, the VSHF energy grows approximately exponentially in time prior to the establishment of equilibrium (a) and the equilibrium VSHF energy increases abruptly as $\varepsilon$ is increased beyond a critical threshold value (b).} \label{fig:modelcomparison} 
\end{figure}

Figure \ref{fig:modelcomparison} (a) shows the time evolution of the kinetic energy of the VSHF and the total energy of the perturbations (the darkest curves show the behavior of the NL system). 
The perturbation kinetic, potential, and total energies are defined as 
\begin{align}
K' = [u'^2+w'^2]/2, && V' = [b'^2]/2N_0^2, && E'=K'+V', \label{eq:perturbation_energies}
\end{align}
where square brackets indicate the domain average. 
The kinetic energy of the VSHF, the potential energy of the buoyancy layers, and the total energy of the horizontal mean state are defined as
\begin{align}
\overline{K} = [U^2]/2, && \overline{V} = [B^2]/2N_0^2, && \overline{E}=\overline{K}+\overline{V}.
\end{align}
In the absence of excitation and dissipation the total energy, $E=\overline{E}+E'$, is conserved. 
The VSHF energy grows approximately exponentially in time before approaching a quasi-steady state in which the VSHF is energetically dominant, weakly fluctuating, and slowly varying in association with the slow variations in the structure of $U$ in Figure \ref{fig:NLphenom} (b). 
Figure \ref{fig:modelcomparison} (b) shows how the VSHF and perturbation energies depend on the nondimensional excitation strength, $\varepsilon k_e^2/r^3$, illustrated using the index zmf defined as \begin{equation}
\text{zmf} \equiv \overline{K}/E,
\end{equation}
which gives the fraction of the total energy that is contained in the VSHF. 
The zmf is small for $\varepsilon\lesssim 50$, with nearly all the energy being contained in the perturbation field. 
As $\varepsilon$ is increased beyond this threshold value, the zmf increases sharply.
This transition in behavior coincides with the emergence of a coherent VSHF.

The abrupt emergence of the VSHF as the excitation strength is increased results from a bifurcation associated with the growth rate of the VSHF-forming instability crossing zero toward positive values at a critical excitation strength \citep{Fitzgerald:2016ux}.  
This bifurcation is predicted by SSD and is reflected in the NL system as shown in Figure \ref{fig:modelcomparison} (b). 


\subsection{QL System}

Before developing the S3T implementation of SSD that we apply in this work it is useful to first develop the QL system.
The QL system is an approximation to the NL system in which the EENL terms are discarded from the perturbation dynamics (\ref{eq:NL1})-(\ref{eq:NL2}) so that nonlinearity is confined to wave-mean flow interactions. 
The QL perturbation dynamics are
\begin{align}
\partial_t \zeta' &= - U\partial_x \zeta' +\partial^2_{zz}U\partial_x \psi' +\partial_x b'+\sqrt{\varepsilon} \xi^{\zeta} -r\zeta'  + \nu \Delta \zeta', \label{eq:QL1} \\
\partial_t b' &= - U\partial_x b' -(N_0^2+\partial_z B)\partial_x \psi' + \sqrt{\varepsilon} \xi^b -rb' + \nu \Delta b'. \label{eq:QL2}
\end{align}
Perturbation equations (\ref{eq:QL1})-(\ref{eq:QL2}) are then coupled to the mean equations (\ref{eq:NL3})-(\ref{eq:NL4}) to produce the closed QL system.  

The QL system incorporates a hypothesis about which aspects of the dynamics are essential to determining the statistical mean equilibrium state of the turbulence, including the large scale structure, and which are inessential. 
In particular, to the extent that wave-mean flow interactions that are spectrally nonlocal are the primary drivers of VSHF formation, the QL system should capture the behavior of the NL system in the VSHF-forming regime. 
Conversely, if arrest of an upscale turbulent cascade at the VSHF scale were mechanistically responsible for the formation of VSHFs then there would be no agreement between NL simulations and QL simulations because the nonlinear interaction among perturbations has been eliminated in QL.
The dark grey curves in Figure \ref{fig:modelcomparison} compare the behavior of the QL system to that of the NL system for the chosen example case. 
The QL system shows good agreement with the NL system, indicating that the dynamical approximations underlying the QL system retain the mechanism responsible for VSHF emergence. 
That VSHF formation does not result from a traditional spectrally local inverse cascade has previously been noted by \citet{Smith:2002wg}.

\subsection{S3T System} \label{sec:S3Tsystem}

The S3T system is a turbulence closure at second order and so necessarily has underlying dynamics that are QL \citep{Herring:1963gx}. 
The analytical simplicity of S3T results from making the ergodic assumption that the horizontal average, which is the appropriate choice of mean for the purpose of analyzing VSHF dynamics, is equivalent to the ensemble average over realizations of the stochastic excitation. 
This ergodic assumption allows the dynamics of the second cumulant to be expressed in the analytical form of a time-dependent Lyapunov equation. 
The ergodic assumption is justified when the domain has sufficient horizontal extent to permit many approximately independent perturbation structures, such as in the case of Figure \ref{fig:NLphenom} (a) in which several perturbation features are visible at each height. 

A derivation of the S3T system in differential form is provided in Section \ref{sec:S3Tderivation} following \citet{Srinivasan:2012im}. 
This approach is complementary to the conventional matrix approach of \citet{Farrell:2003ud}. 
The continuous approach is useful for carrying out linear stability analysis and for deriving closed-form dispersion relations for instability growth rates that are amenable to asymptotic analysis. 
The matrix approach is required when performing S3T analysis of the finite-amplitude structure and equilibration dynamics of the VSHF following its initial emergence. 
A derivation of the S3T system following the matrix approach can be found in \citet{Fitzgerald:2016ux}.

The light grey curves in Figure \ref{fig:modelcomparison} show the behavior of the matrix S3T system. 
In S3T, the perturbation fields and the stochastic excitation are described by their covariance matrices. 
The S3T integrations shown in Figure \ref{fig:modelcomparison} use an excitation covariance matrix corresponding to the ring excitation used in the NL and QL simulations. 
The S3T system was integrated numerically to equilibrium using a fourth-order Runge-Kutta method with resolution of 128 grid points in the vertical direction and eight Fourier components in the horizontal direction. 
To allow for comparison to be made between the time evolution of the S3T system and the NL and QL systems the S3T integration shown in panel (a) was initialized using the mean fields and instantaneous perturbation covariance matrix diagnosed from the QL simulation at $t=5$. 
In the S3T integrations shown in panel (b) the perturbation covariance matrix was instead initialized to correspond to the homogeneous turbulence fixed point given by (\ref{eq:homoFP}) and the mean fields were initialized as small random perturbations. 

The VSHF emergence diagnostics in S3T are in good general agreement with those of the NL and QL systems. 
However, we note that S3T exhibits an exact bifurcation structure in which the zmf is exactly zero below the critical excitation strength and sharply increases beyond it, whereas the NL and QL zmfs have small but nonzero values for excitations less than the critical excitation that corresponds to the bifurcation point (Figure \ref{fig:modelcomparison} (b)). 
The nonzero values of the zmf in NL and QL for excitations less than that of the bifurcation point result from the excitation of weakly damped VSHF modes by the stochastic fluctuations in those systems \citep{Constantinou:2014fh}. 
The vertical dashed line in Figure \ref{fig:modelcomparison} (b) shows the VSHF bifurcation point predicted by the S3T dispersion relation derived in Section \ref{sec:dispersion}. 
This prediction is in good agreement with the results of the NL and QL systems and corresponds exactly with the behavior of the matrix S3T system. 
We also note that, although the underlying dynamics of the S3T system are QL, the time average of the QL system in statistical equilibrium does not exactly equal the fixed point equilibrium state of the S3T system. 
Although these states are often similar, the QL system formally converges to the S3T system in the limit that the perturbation covariances are calculated from an infinite ensemble of realizations of (\ref{eq:QL1}) and (\ref{eq:QL2}), rather than in the limit of an infinite time average \citep{Farrell:2003ud}. 

\section{S3T Equations of Motion} \label{sec:S3Tderivation}

We now develop the S3T equations of motion following the differential approach of \citet{Srinivasan:2012im}. 
The dynamical variables characterizing the perturbation field in S3T are the two-point equal-time ensemble mean perturbation covariance functions. 
For example, the vorticity covariance is defined as 
\begin{equation}
Z(\vec{x}_1,\vec{x}_2,t)\equiv \langle \zeta'_1(t)\zeta'_2(t)\rangle,
\end{equation}
and the other required covariances are
\begin{align}
\Psi &\equiv \langle \psi'_1 \psi'_2 \rangle, & T &\equiv \langle b_1' b_2' \rangle,  \\
\Gamma^{\zeta} &\equiv \langle \zeta'_1 b'_2 \rangle, & \Gamma^b &\equiv \langle b_1' \zeta'_2 \rangle, \\
S^{\zeta} &\equiv \langle \psi'_1 b'_2 \rangle, & S^b &\equiv \langle b_1' \psi'_2 \rangle.
\end{align}
The covariances $\Gamma^{\zeta,b}$ and $S^{\zeta,b}$ are related through $\Gamma^{\zeta}=\Delta_1 S^{\zeta}$ and $\Gamma^b=\Delta_2 S^b$, where $\Delta_i \equiv \partial^2/\partial x_i^2+ \partial^2/\partial z_i^2$. 
We define them separately for convenience. 
Similar to equation (\ref{eq:GzetaGb_exchange}) for the excitation cross-covariances, $G^{\zeta}$ and $G^b$, the perturbation cross-covariances, $\Gamma^{\zeta,b}$ and $S^{\zeta,b}$, obey the symmetry relations
\begin{align}
\Gamma^{\zeta}(\vec{x}_1,\vec{x}_2) = \Gamma^b(\vec{x}_2,\vec{x}_1), && S^{\zeta}(\vec{x}_1,\vec{x}_2)=S^b(\vec{x}_2,\vec{x}_1). \label{eq:GammaSexchange}
\end{align}

Equations of motion for $Z$, $T$, and $\Gamma^{\zeta}$ can be derived straightforwardly from the QL system using the ergodic approximation. 
We express the dynamics using the collective coordinates
\begin{align}
x&=x_1 - x_2, && z=z_1-z_2, \\
\bar{x} &= (x_1+x_2)/2, && \bar{z}=(z_1+z_2)/2.
\end{align}
Assuming that the turbulence is statistically homogeneous in the horizontal direction, as suggested by Figure \ref{fig:NLphenom} (a), we take all covariances to be independent of $\bar{x}$. 
Covariances may, however, depend on $\bar{z}$ because the emergent vertical banding breaks the homogeneity of the turbulence in the vertical direction. 
Under the assumption of horizontal homogeneity the operators $\partial_{z,i}$ and $\Delta_i$ can be written as 
\begin{align}
\partial_{z,i} =(-1)^{1+i}\partial_z+\frac12 \partial_{\bar{z}}, \label{eq:partialderivs} && \Delta_i  = \Delta - (-1)^i \partial^2_{z\bar{z}} + \frac14 \partial^2_{\bar{z}\bar{z}},
\end{align}
in which $\Delta$ is the Laplacian in the difference variables, $\Delta=\partial^2_{xx}+\partial^2_{zz}$. 

The covariance dynamics are
\begin{widetext}
\begin{align}
\partial_t Z +(U_1-U_2)\partial_x Z + (U_1''+U_2'')\partial^3_{xz\bar{z}}\Psi -(U_1''-U_2'')(\Delta+\frac14 \partial^2_{\bar{z}\bar{z}})\partial_x \Psi &= -2rZ+\partial_x(\Gamma^b-\Gamma^{\zeta})+\varepsilon \Xi, \label{eq:S3T1} \\
\partial_t T + (U_1-U_2)\partial_x T + N_0^2 \partial_x (S^{\zeta}-S^b)+B_1'\partial_x S^{\zeta}-B_2'\partial_x S^b &= -2rT+\varepsilon \Theta, \label{eq:S3T2} \\
\partial_t \Gamma^{\zeta} + (U_1-U_2)\partial_x \Gamma^{\zeta}-U_1''\partial_x S^{\zeta} -(N_0^2+B'_2)(\Delta+\partial^2_{z\bar{z}}+\frac14 \partial^2_{\bar{z}\bar{z}})\partial_x \Psi &= -2r\Gamma^{\zeta}+\partial_x T + \varepsilon G^{\zeta}, \label{eq:S3T3}
\end{align}
\end{widetext}
%
%
%
%
%
%
%
%
\hspace{-4mm} where $U_i''$ denotes the curvature of $U$ at $z_i$. 
The equation of motion for $\Gamma^b$ can be obtained from (\ref{eq:S3T3}) using (\ref{eq:GammaSexchange}). 
Viscous terms included to ensure numerical convergence in NL and QL simulations are excluded from the present development for simplicity but can be straightforwardly included. 

To obtain a closed dynamics, the equations of motion for $U$ and $B$ must also be expressed in terms of the perturbation covariances. 
The turbulent momentum and buoyancy fluxes can be written as 
\begin{align}
\langle u'w' \rangle = \partial^2_{xz} \Psi \Big|_{x=z=0}, &&
\langle w'b' \rangle = \frac12 \partial_x (S^{\zeta}-S^b)\Big|_{x=z=0}.  \label{eq:eddymomentumflux} 
\end{align}
The dynamics of $U$ and $B$ then become
\begin{align}
\partial_t U &= -r_m U - \partial^3_{xz\bar{z}} \Psi \Big|_{x=z=0}, \label{eq:S3T4} \\ 
\partial_t B &= -r_m B - \frac12 \partial^2_{x\bar{z}} (S^{\zeta}-S^b)\Big|_{x=z=0}. \label{eq:S3T5}
\end{align}
Equations (\ref{eq:S3T1})-(\ref{eq:S3T3}) together with (\ref{eq:S3T4})-(\ref{eq:S3T5}) constitute the closed S3T system. 
Technical details useful for the derivation of (\ref{eq:S3T1})-(\ref{eq:S3T5}) can be found in \citet{Srinivasan:2012im}. 

Before proceeding to the analysis of S3T it is useful to express aspects of the energetics in terms of covariances. 
The ensemble mean values of $K'$ and $V'$ are given by 
\begin{align}
\langle K' \rangle =-\frac12 \left[(\Delta-\frac14 \partial^2_{\bar{z}\bar{z}})\Psi \right]_{x=z=0}, && \langle V' \rangle = \frac12 N_0^{-2} \left[T\right]_{x=z=0},
\end{align}
where square brackets indicate the average over $\bar{z}$. 
From (\ref{eq:S3T1})-(\ref{eq:S3T2}), the rates at which kinetic and potential energy are injected into the perturbation field by the stochastic excitation are
\begin{align}
\varepsilon_K=-\frac{\varepsilon}{2}\Delta^{-1}\Xi \Big|_{x=z=0}, && \varepsilon_V = \frac{\varepsilon}{2N_0^2} \Theta \Big|_{x=z=0}.
\end{align}
The excitation strength control parameter, $\varepsilon$, and the excitation structure functions, $\Xi$ and $\Theta$, collectively determine the overall amplitude of the excitation, its spatial structure, and how the injected energy is partitioned between kinetic and potential forms. 
We choose the convention that the functions $\Xi$ and $\Theta$ set the spatial structure of the excitation and the ratio $\varepsilon_K/\varepsilon_V$, while the control parameter $\varepsilon$ scales the total energy injection rate, $\varepsilon_K+\varepsilon_V$. 
We further choose to normalize the functions $\Xi$ and $\Theta$ such that the total energy injection rate is equal to the value of the parameter $\varepsilon$, so that $\varepsilon_K+\varepsilon_V=\varepsilon$. 

It is useful to express $\varepsilon_K$ and $\varepsilon_V$ in terms of the Fourier transforms of the excitation covariances. 
Using the Fourier conventions 
\begin{align}
f(\vec{x}) = \iint \frac{\mbox{d}p \mbox{d}q}{(2\pi)^2} \tilde{f}(\vec{p})e^{i\vec{p}\cdot \vec{x}}, && \tilde{f}(\vec{p}) = \iint \mbox{d}x \mbox{d}z f(\vec{x})e^{-i\vec{p}\cdot\vec{x}}, \label{eq:Fourier}
\end{align}
with $\vec{p}=(p,q)$ and $h^2=p^2+q^2$, we have
\begin{align}
\varepsilon_K &= \varepsilon \iint \frac{\mbox{d}p \mbox{d}q}{(2\pi)^2} \frac{\tilde{\Xi}}{2h^2}\equiv \varepsilon \iint \frac{\mbox{d}p \mbox{d}q}{(2\pi)^2} \tilde{K}(p,q), \\
\varepsilon_V &= \varepsilon \iint \frac{\mbox{d}p \mbox{d}q}{(2\pi)^2} \frac{\tilde{\Theta}}{2 N_0^2}\equiv \varepsilon \iint \frac{\mbox{d}p \mbox{d}q}{(2\pi)^2} \tilde{V}(p,q), \\
\varepsilon &= \varepsilon_K+\varepsilon_V \equiv \varepsilon \iint \frac{\mbox{d}p \mbox{d}q}{(2\pi)^2} \tilde{E}(p,q),  \label{eq:varepsilon_spectral}
\end{align}
where we have defined the functions $\tilde{K}=\tilde{\Xi}/(2h^2)$, $\tilde{V}=\tilde{\Theta}/(2N_0^2)$, and $\tilde{E}=\tilde{K}+\tilde{V}$ which characterize the spectral structures of the kinetic, potential, and total energy injection rates. 
In our normalization the integral in (\ref{eq:varepsilon_spectral}) is equal to 1, as $\tilde{E}$ controls the spectral distribution of the excitation but not its total energy injection rate. 

\section{S3T Stability of Homogeneous Stratified Turbulence} \label{sec:linearization}

We next apply S3T to analyze the possibility of emergent vertical banding such as that observed in Figure \ref{fig:NLphenom}. 
We begin by considering the alternate possibility that no coherent structures exist and that the turbulence is statistically steady and homogeneous. 
S3T admits a fixed point solution corresponding to such a homogeneous state. 
We analyze the linear stability of this solution to determine the rates of growth or decay of perturbations to homogeneous turbulence associated with VSHFs and horizontal mean buoyancy layers. 
If perturbations with positive growth rates exist, the underlying homogeneous turbulence is unstable to the development of vertical banding, which provides an explanation for the initial emergence of structure as observed in simulations. 

Statistically steady homogeneous turbulence is characterized by $U=B=\partial_t=\partial_{\bar{z}}=0$. 
Homogeneous S3T equilibria, whose covariance functions are indicated by a subscript $H$, obey the linear equations
\begin{align}
0 &= -2rZ_H+\partial_x(\Gamma^b_H-\Gamma^{\zeta}_H)+\varepsilon \Xi, \label{eq:S3Thomo1} \\
0 &= -N_0^2 \partial_x (S^{\zeta}_H-S^b_H)-2rT_H+\varepsilon \Theta, \label{eq:S3Thomo2} \\
0 &= N_0^2 \partial_x \Delta \Psi_H-2r\Gamma^{\zeta}_H+\partial_x T_H + \varepsilon G^{\zeta}. \label{eq:S3Thomo3}
\end{align}
Equations (\ref{eq:S3Thomo1})-(\ref{eq:S3Thomo3}) can be solved for general $\Xi$, $\Theta$, and $G^{\zeta}$. 
However, a great simplification occurs when the excitation is chosen such that the potential and kinetic energy injection rates are equal at each wavenumber and also such that the excitations of the vorticity and buoyancy fields are uncorrelated. 
These conditions correspond to the relations 
\begin{align}
\tilde{K}=\tilde{V}=\frac{1}{2}\tilde{E}, && G^{\zeta}=G^b=0. \label{eq:ExcitationAssumption}
\end{align}
The excitation structures shown in Figure \ref{fig:excitation_phys} have these fairly natural properties. 
When (\ref{eq:ExcitationAssumption}) holds, the solution of (\ref{eq:S3Thomo1})-(\ref{eq:S3Thomo3}) is given by
\begin{align}
Z_H=\frac{\varepsilon \Xi}{2 r}, && T_H=\frac{\varepsilon \Theta}{2r}, && \Gamma^{\zeta,b}_H=S^{\zeta,b}_H=0. \label{eq:homoFP}
\end{align}
We note that the fixed point solution (\ref{eq:homoFP}) does not depend on $N_0^2$. 
An analogous result has been obtained for $\beta$-plane turbulence, in which the homogeneous turbulent state does not depend on $\beta$ \citep{Srinivasan:2012im}.

We now outline the linear stability analysis of the fixed point (\ref{eq:homoFP}). 
Details are provided in Appendix A. 
We begin by expanding the perturbation covariances to first-order about the fixed point as $Z(x,z,\bar{z},t)=Z_{H}(x,z)+\delta Z(x,z,\bar{z},t)$ and similarly for $T$, $\Gamma^{\zeta}$, and $\Gamma^b$. 
Horizontal mean structures are expanded about zero as $U(\bar{z},t)=\delta U(\bar{z},t)$ and $B(\bar{z},t)=\delta B(\bar{z},t)$. 
Substitution into (\ref{eq:S3T1})-(\ref{eq:S3T5}) then yields a set of linearized equations such as
\begin{multline}
\partial_t \delta Z + (\delta U _1 -\delta U _2 )\partial_x Z_H - (\delta U_1''-\delta U_2 '' )\Delta \partial_x \Psi_H \\
-\partial_x(\delta \Gamma^b-\delta \Gamma^{\zeta})=-2r\delta Z,  \label{eq:S3T_lin_phys_1_MAIN} 
\end{multline}
which governs the perturbation to the vorticity covariance. 
It is then useful to write the perturbation variables in the Fourier ansatz
\begin{align}
\delta C(x,z,\bar{z},t) &= e^{st}e^{im\bar{z}} \hat{C}(x,z)_{m,s}, \label{eq:FourierAnsatz1} \\
\delta U(\bar{z},t) &= e^{st}e^{im\bar{z}} \hat{U}_{m,s}, \label{eq:FourierAnsatz2} \\
\delta B(\bar{z},t) &= e^{st}e^{im\bar{z}} \hat{B}_{m,s}, \label{eq:FourierAnsatz3}
\end{align}
plus complex conjugate terms, where $C$ is a placeholder for $Z$, $T$, $\Gamma^{\zeta}$, and $\Gamma^b$. 
The perturbation covariance coefficients, $\hat{C}(x,z)$, are further decomposed using their Fourier transforms as in (\ref{eq:Fourier}). 
Expressing the perturbation equations using these Fourier representations for the statistical variables after some manipulation we obtain an eigenproblem for the eigenvalue, $s$, whose real part is equal to the growth rate of banded perturbations with vertical wavenumber $m$ in homogeneous turbulence.

The eigenproblem simplifies dramatically when the excitation structure is chosen to have the reflection symmetry $\tilde{E}(p,q)=\tilde{E}(-p,q)$, which corresponds to equal excitation of gravity wave modes with positive and negative phase speeds. 
This property is typical of excitation structures chosen in theoretical studies of stratified turbulence and is possessed by the excitation structures shown in Figure \ref{fig:excitation_phys}.
For reflection-symmetric excitation, the eigenproblem for $s$ factors into two decoupled eigenproblems, each of which determines a set of modes and their growth rates. 
The first set of modes, which we call the VSHF modes, has $\delta B = 0$ so that the horizontal mean structure consists purely of a VSHF perturbation with no buoyancy layer perturbation. 
The second set of modes, which we call the buoyancy layer modes, has $\delta U=0$ so that the horizontal mean structure is a pure buoyancy layer perturbation. 
Each set of modes has its own dispersion relation. 
Denoting by $s_U$ the eigenvalues corresponding to the VSHF modes and by $s_B$ the eigenvalues corresponding to the buoyancy layer modes, the dispersion relations are
\begin{align}
\frac{\bar{s}_U}{s'_U}= \varepsilon \int \text{d}p\text{d}q \mathcal{F}_U(p,q,m,N_0^2,r,s_U)\tilde{E}(p,q), \label{eq:dispersion_U} \\
\frac{\bar{s}_B}{s'_B}= \varepsilon \int \text{d}p\text{d}q \mathcal{F}_B(p,q,m,N_0^2,r,s_B)\tilde{E}(p,q), \label{eq:dispersion_B}
\end{align}
where $\bar{s}_{U,B}=s_{U,B}+r_m$ and $s'_{U,B}=s_{U,B}+2r$. 
$\mathcal{F}_U$ and $\mathcal{F}_B$ are functions whose detailed forms are provided in Appendix A. 
Equations (\ref{eq:dispersion_U})-(\ref{eq:dispersion_B}) are resistant to analytical solution because the eigenvalues, $s_U$ and $s_B$, appear in the integrands. 
However, the eigenvalues can be computed numerically and approximate analytical solutions can be obtained in a variety of cases. 
In the next section we apply (\ref{eq:dispersion_U})-(\ref{eq:dispersion_B}) to analyze the stability of homogeneous turbulence maintained by the IRE and MCE structures discussed in Section \ref{sec:NLphenom}. 

\section{Application of the Dispersion Relations to the Cases of IRE and MCE} \label{sec:dispersion}

\subsection{Isotropic Ring Excitation (IRE)} \label{sec:dispersionIRE}

We first analyze the VSHF-forming and buoyancy layering instabilities in stratified turbulence maintained by IRE. 
Nondimensionalizing length by the excitation scale, $1/k_e$, and time by the perturbation damping time, $1/r$, recalling the definition $h^2=p^2+q^2$ the energy injection spectrum is given by 
\begin{equation}
\tilde{E}_{IRE}(p,q) = 2\pi \delta (h-1). \label{eq:varepsIRE}
\end{equation}
Figure \ref{fig:excitation_phys} (a) shows a realization in physical space of an excitation approximating (\ref{eq:varepsIRE}). 
The IRE dispersion relations are obtained by evaluating (\ref{eq:dispersion_U})-(\ref{eq:dispersion_B}) for $\tilde{E}=\tilde{E}_{IRE}$ (see Appendix B).
Figure \ref{fig:s_vs_n} (a) shows how $s_U$ and $s_B$ vary with $m$, the vertical wavenumber of the emergent banding, for several representative $N_0^2$ values\footnote{We note that for the parameters considered in Figure \ref{fig:s_vs_n}, $s_U$ and $s_B$ are real. 
Although the eigenvalues can be complex under some circumstances, we have found through experimentation that these cases do not typically correspond to the most unstable modes and as such we treat the case of real eigenvalues throughout this paper.}. 
The parameters used are $\varepsilon=75$, chosen so that $s_U>0$ over some band of $m$ for each selected value of $N_0^2$, and  $r_m=0.1$, chosen to make contact with the simulations shown in Figures \ref{fig:NLphenom} and \ref{fig:modelcomparison} and with our previous work \citep{Fitzgerald:2016ux}. 


\begin{figure}[t]
\centerline{\includegraphics[scale=1]{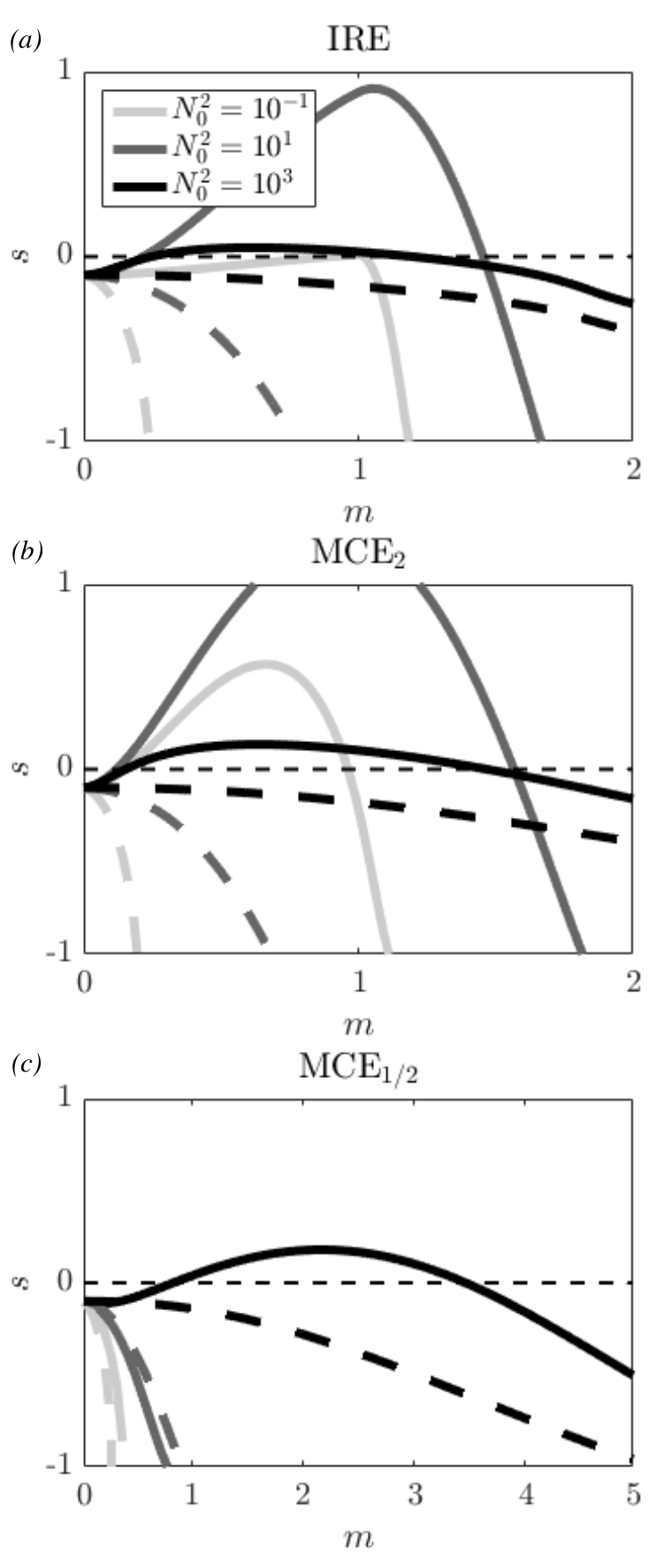}}
\caption{Growth rates of the VSHF-forming (solid) and buoyancy layering (dashed) instabilities as functions of the vertical wavenumber, $m$, of the horizontal mean structure for turbulence excitation structures (a) IRE, (b) MCE$_2$, and (c) MCE$_{1/2}$. Three representative $N_0^2$ values are shown: $N_0^2=10^{-1}$ (weak stratification, light curves), $N_0^2=10^1$ (intermediate stratification, medium curves), and $N_0^2=10^3$ (strong stratification, darkest curves). The parameters used are $\varepsilon=75$, $r_m=0.1$, and $\nu=0$.} \label{fig:s_vs_n} 
\end{figure}


The properties of the VSHF-forming instability depend on the stratification. 
Setting the notation $s_U^{\star}=\text{max}_m[s_U(m)]=s_U(m^{\star})$, 
under weak stratification (Figure \ref{fig:s_vs_n} (a), light solid curve) the fastest-growing VSHF corresponds to $m^{\star}\approx1$, and $s_U^{\star}$ is only weakly positive. 
VSHFs with $0<m\lesssim1$ have negative growth rates but decay more slowly than the explicit VSHF damping rate, \emph{i.e.,} $-r_m<s_U<0$. 
This indicates that relatively large-scale VSHFs are reinforced by IRE turbulence. 
VSHFs with $m\gtrsim1$ have $s_U<-r_m$, indicating that relatively small-scale VSHFs are dissipated by IRE turbulence.

\begin{figure}[t]
\centerline{\includegraphics[scale=1]{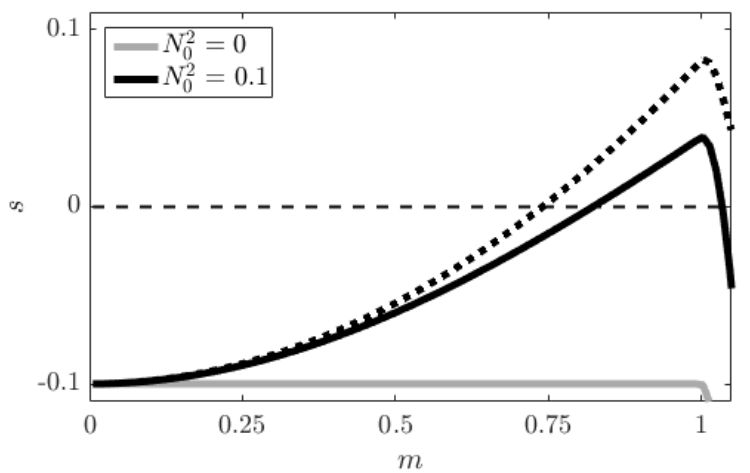}}
\caption{Growth rate of the VSHF-forming instability as a function of $m$ under zero and weak stratification in the case of IRE. The light grey curve shows $s_U(N_0^2=0)$ and the solid black curve shows how $s_U$ is enhanced when weak stratification is introduced. The thick dashed curve shows the asymptotic approximation (\ref{eq:IREsmallNsqsU}). The parameters used are $\varepsilon=50$, $r_m=0.1$ and $\nu=0$. } \label{fig:SmallNsqAsymp} 
\end{figure}

For small $N_0^2$, the VSHF-forming instability can be understood by perturbing about the case of unstratified IRE turbulence (see Appendix B). 
For $N_0^2=0$, the induced momentum flux, $\langle u'w'\rangle$, vanishes when the VSHF scale is larger than the excitation scale, so that $s_U=-r_m$ for $0\le m \le 1$. 
For perturbatively weak stratification, $s_U$ is modified to 
\begin{equation}
s_U \approx -r_m + \frac{ \varepsilon m^2 N_0^2}{8(2-r_m)^3}, \label{eq:IREsmallNsqsU}
\end{equation}
which is valid for $0<m<1$. 
The estimate (\ref{eq:IREsmallNsqsU}) is compared to the result from (\ref{eq:dispersion_U}), and to the $N_0^2=0$ solution, in Figure \ref{fig:SmallNsqAsymp}. 
Quadratic enhancement of $s_U$ as a function of $m$ leads to $m^{\star}=1$ for weak stratification. 
Figure \ref{fig:s_vs_n} (a) shows that $m^{\star}\approx1$ also corresponds to the fastest-growing VSHF at the intermediate stratification $N_0^2=10$. 
For $m>1$ equation (\ref{eq:IREsmallNsqsU}) is replaced by a more complex expression but the details are inessential due to the strong dissipation of these VSHFs by the zeroth order unstratified turbulence. 


The quadratic increase of $s_U$ with $m$ in (\ref{eq:IREsmallNsqsU}) suggests that the VSHF-forming instability results from negative eddy viscosity.
Indeed, for $m\ll 1$, (\ref{eq:dispersion_U}) gives
\begin{align}
s_U \approx& -r_m - \nu_{eddy}m^2, \label{eq:eddy_viscosity_1}\\
\nu_{eddy} &= -\varepsilon g(N_0^2,r_m), \label{eq:eddy_viscosity_2}
\end{align}
where $g$ is a positive-definite function (see Appendix B) so that $\nu_{eddy}<0$ for all $N_0^2$.
We analyze the dynamics leading to $\nu_{eddy}<0$ in Sections \ref{sec:feedback} and \ref{sec:processes}. 
In IRE $\beta$-plane turbulence, large-scale jets instead initially form due to negative eddy hyperviscosity \citep{Srinivasan:2012im}. 

\begin{figure}[t]
\centerline{\includegraphics[scale=1]{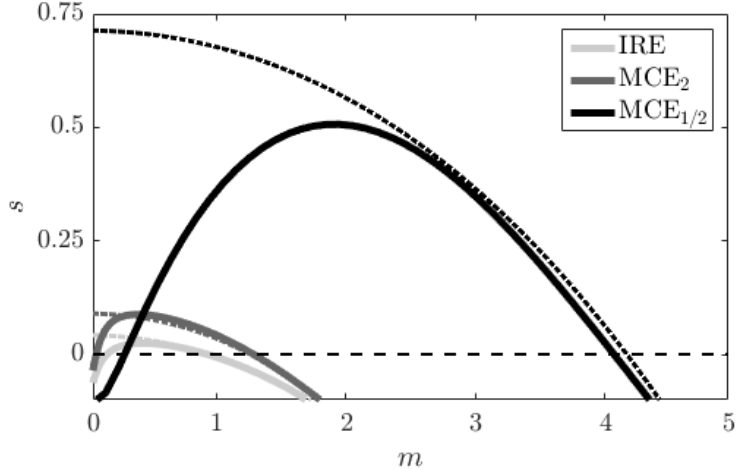}}
\caption{Growth rates of the VSHF-forming instability under strong stratification for the IRE, MCE$_2$, and MCE$_{1/2}$ cases. Solid curves show growth rates calculated using the full dispersion relation and dashed curves show asymptotic approximations (see Appendices D and E). The structure of $s_U$ is generic, with the fastest growing wavenumber approaching $m=0$ as $N_0^2\to\infty$. The parameters used are $\varepsilon=50$, $N_0^2=10^5$, $r_m=0.1$, and $\nu=0$.} \label{fig:LargeNsqAsymp} 
\end{figure}

Both $s_U^{\star}$ and $m^{\star}$ are significantly modified as $N_0^2$ is increased. 
The value of $s_U^{\star}$ increases to a maximum near $N_0^2=10$ (Figure \ref{fig:s_vs_n} (a)). 
As the stratification becomes strong, (\ref{eq:dispersion_U}) gives (see Appendix B)
\begin{equation}
s_U \approx -r_m + (\varepsilon/N_0^2)(1-r_m/2)(3-m^2), \label{eq:IRE_LargeNsq_sU}
\end{equation}
which is compared to the unapproximated result in Figure \ref{fig:LargeNsqAsymp} (light grey curve) for $N_0^2=10^5$ . 
The behavior of $s_U$ is captured by (\ref{eq:IRE_LargeNsq_sU}) when $m$ is not small. 
When $m$ is small, the behavior of $s_U$ is instead captured by (\ref{eq:eddy_viscosity_1}). 
As $N_0^2$ increases, (\ref{eq:IRE_LargeNsq_sU}) indicates that the growth rate weakens ($s_U^{\star}\to-r_m$) and the VSHF emerges at larger and larger scale ($m^{\star}\to0$). 
Similar results are found for zonostrophic instability, with the jet growth rate weakening and the jet scale increasing as $\beta\to\infty$. 
This behavior is attributed to disruption of wave-mean flow interaction between travelling waves and stationary jets by the increase in Rossby wave group velocity \citep{Bakas:2015iy}. 
This mechanism likely operates in stratified turbulence as well, with the increasing group velocity of gravity waves at large $N_0^2$ disrupting the wave-mean flow interaction underlying VSHF formation. 


The dashed curves in Figure \ref{fig:s_vs_n} (a) show how the growth rate of the buoyancy layering instability, $s_B$, varies with $m$. 
In all cases shown, $s_B<-r_m$, indicating that perturbations to homogeneous turbulence associated with buoyancy layers are dissipated by downgradient eddy buoyancy fluxes. 
Although this result appears to contradict the results of the NL simulation shown in Section \ref{sec:NLphenom}, which forms buoyancy layers, our previous work has shown that these buoyancy layers do not emerge in homogeneous turbulence but instead form nonlinearly once a finite-amplitude VSHF has emerged \citep{Fitzgerald:2016ux}.

\subsection{Monochromatic Excitation (MCE)} \label{sec:dispersionMCE}

VSHF formation occurs in stratified turbulence for a wide range of parameter choices and excitation structures, but the properties of the VSHF-forming instability can depend on the properties of the excitation.
To demonstrate this we analyze the VSHF-forming instability in turbulence maintained by MCE, which is a red-noise structure that differs qualitatively from IRE while retaining comparable analytical simplicity. 
MCE excites a single horizontal wavenumber component, $k_0$, with a Gaussian energy injection spectrum in vertical wavenumber. 
Nondimensionalizing\footnote{Although our nondimensionalizations for the IRE and MCE cases differ slightly, results in the various cases can be directly compared for the same parameters when the ring wavenumber, $k_e$, is set equal to the MCE wavenumber, $k_0$.} length by the horizontal scale of the excitation, $1/k_0$, and time by the perturbation damping time,  $1/r$, the MCE energy injection spectrum is given by
\begin{equation}
\tilde{E}_{MCE}(p,q) = \pi^{3/2}\ell_c \exp (-\ell_c^2 q^2/4) \left[\delta(p+1)+\delta(p-1) \right]. \label{eq:varepsMCE}
\end{equation}
The parameter $\ell_c$ is equal to the correlation length of the excitation in the vertical direction and sets the width of the spectrum in the vertical wavenumber, $q$. 
We analyze the cases MCE$_2$, with $\ell_c=2$, and MCE$_{1/2}$, with $\ell_c=1/2$, and compare them to the case of IRE. 
Physical-space realizations of MCE$_2$ and MCE$_{1/2}$ are shown in Figures \ref{fig:excitation_phys} (b) and (c). 
MCE$_2$ excites structures with somewhat greater vertical extent than horizontal extent, while MCE$_{1/2}$ excites structures with comparable or smaller vertical extent than horizontal extent. 

The dispersion relations of the VSHF-forming and buoyancy layering instabilities for MCE are obtained by evaluating (\ref{eq:dispersion_U})-(\ref{eq:dispersion_B}) with $\tilde{E}=\tilde{E}_{MCE}$ (see Appendix C). 
The growth rates, $s_U$ and $s_B$, are shown in Figure \ref{fig:s_vs_n} (b) and (c) for MCE$_2$ and MCE$_{1/2}$ as functions of the vertical wavenumber, $m$, of the emergent banding. 

For small $N_0^2$ (light grey curves in Figure \ref{fig:s_vs_n}) the properties of the VSHF-forming instability differ markedly among the three excitation structures. 
Although the instability is weak for IRE as expressed by  (\ref{eq:IREsmallNsqsU}), $s_U$ remains positive as $N_0^2\to0$ for MCE$_2$ and remains strongly negative as $N_0^2\to0$ for MCE$_{1/2}$.
These results are consistent with previous work on  unstratified turbulence by \citet{Bakas:2011bt} using a slightly modified formulation of MCE. 

As $N_0^2$ is increased, the IRE and MCE$_2$ cases behave similarly, with $s_U^{\star}$ increasing to a maximum near $N_0^2=10$ before decaying like $1/N_0^2$. 
The MCE$_{1/2}$ case behaves quite differently, with $s_U^{\star}<0$ for weak and intermediate stratification values and $s_U^{\star}>0$ first occurring for $N_0^2\approx 10^2$. 
Figure \ref{fig:LargeNsqAsymp} compares $s_U$ across the cases for $N_0^2=10^5$. 
The asymptotic behavior of $s_U$ as $N_0^2\to\infty$ is generic among the cases, but the band of wavenumbers for which the VSHF is supported by the eddy fluxes, as well as the maximum VSHF growth rate, differs between the cases. 
Surprisingly, the MCE$_{1/2}$ case exhibits the largest growth rates among all cases for large $N_0^2$, with the VSHF emerging at the smallest vertical scale. 
We analyze the mechanism responsible for these properties in Sections \ref{sec:feedback} and \ref{sec:processes}. 

The dashed curves in Figures \ref{fig:s_vs_n} (b) and (c) show the growth rate of the buoyancy layering instability, $s_B$, for MCE. 
As was found for IRE, $s_B<0$ in all cases shown. 
The failure of the buoyancy layering mode to obtain positive growth rates for either canonical ring or red-noise excitation suggests that the initial formation of buoyancy layers is unlikely to arise from an instability of the homogeneous turbulence. 

\section{Stability boundaries} \label{sec:stabilityboundary}

\subsection{IRE}
The dependence of the VSHF-forming instability on the control parameters and excitation structure can be concisely displayed using the stability boundary or neutral curve. 
The stability boundary is defined as the critical value of $\varepsilon$, denoted $\varepsilon_c$, at which $s_U$ first becomes positive as $\varepsilon$ is increased.  
When the emergent VSHF is stationary so that $s_U$ is real, the stability boundary coincides with the simultaneous conditions $s_U=0$ and $\partial_m s_U=0$. 
Alternatively, we may obtain the critical excitation strength for each VSHF wavenumber by setting $s_U=0$ in (\ref{eq:dispersion_U}), which gives
\begin{equation}
\varepsilon_c(m)=\frac{r_m}{2r}\left[\int \text{d}p\text{d}q\mathcal{F}_U\Big|_{s_U=0} \tilde{E} \right]^{-1}. \label{eq:stabilityboundary}
\end{equation}
The stability boundary is then given by 
\begin{equation}
\varepsilon_c = \text{min}_{m} [\varepsilon_c(m)]=\varepsilon_c(m^{\star}). \label{eq:varepsc_minimization}
\end{equation}

\begin{figure}[t]
\centerline{\includegraphics[scale=1]{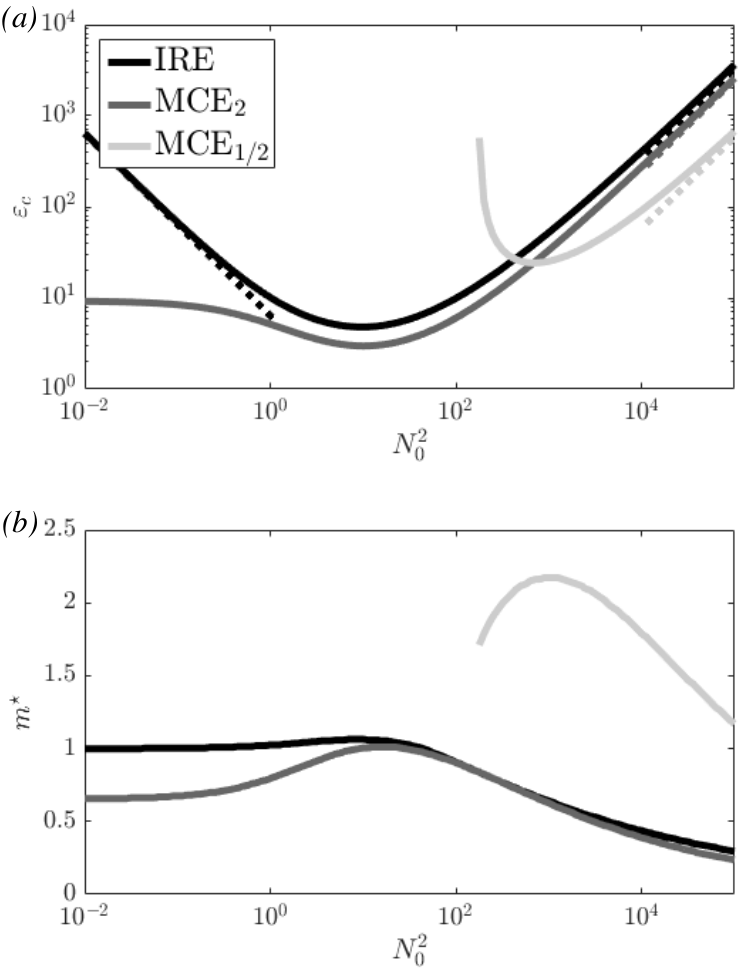}}
\caption{Stability boundary (a) and emergent VSHF wavenumber (b) for the IRE, MCE$_2$, and MCE$_{1/2}$ cases. Solid curves in (a) show the value of $\varepsilon_c$ calculated using full dispersion relation and dashed lines show asymptotic approximations (see Appendices D and E). The parameters used are $r_m=0.1$ and $\nu=0$.} \label{fig:StabilityBoundaryAsymptotics} 
\end{figure}

Figure \ref{fig:StabilityBoundaryAsymptotics} (darkest curves) shows $\varepsilon_c$ (panel a) and the emergent VSHF wavenumber, $m^{\star}$ (panel b), for the case of IRE. 
Dashed lines provide asymptotic approximations (see Appendix B). 
The stability boundary reflects the properties of $s_U$ discussed in Section \ref{sec:dispersion}\ref{sec:dispersionIRE}. 
When the stratification is weak, $\varepsilon_c$ grows like  $1/N_0^2$ and $m^{\star}$ approaches $1$. 
This behavior reflects the small-$N_0^2$ structure of $s_U$ described by (\ref{eq:IREsmallNsqsU}) and shown in Figure \ref{fig:SmallNsqAsymp}. 
Because the VSHF-forming instability develops perturbatively with increasing stratification, very weak stratification requires very strong excitation to produce an instability. 
As $N_0^2$ increases, $\varepsilon_c$ decreases to a minimum near $N_0^2=10$, reflecting the peak in $s_U^{\star}$ near that value of $N_0^2$ visible in Figure \ref{fig:s_vs_n} (a). 
For very strong stratification, $\varepsilon_c$ again becomes large, growing like $N_0^2$ for large $N_0^2$, with $m^{\star}\to0$. 
This behavior reflects the large-$N_0^2$ structure of $s_U$ described by (\ref{eq:IRE_LargeNsq_sU}) and shown in Figure \ref{fig:LargeNsqAsymp}. 


\begin{figure}[t]
\centerline{\includegraphics[scale=1]{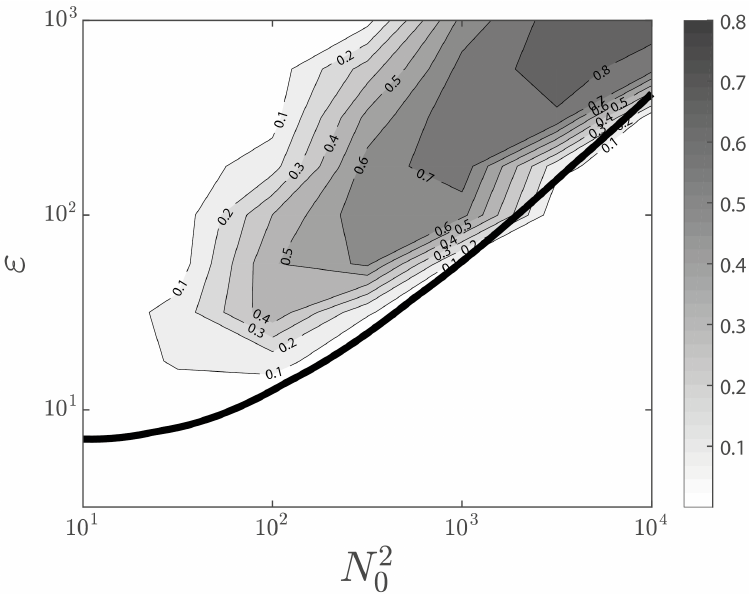}}
\caption{Time average zmf value in NL simulations (shaded contours) compared with the stability boundary of the VSHF-forming instability predicted by S3T (solid curve). Turbulence is maintained with IRE and the time average zmf is calculated following a spin-up period. For intermediate and strong stratification the S3T stability boundary captures the abrupt emergence of the VSHF as $\varepsilon$ is increased, including the increase of the critical excitation strength as $N_0^2$ is increased. The parameters used are $r_m=0.1$ and $\nu=0.03$.} 
\label{fig:NLvsS3T_stabilityboundary}
\end{figure}

We compare the S3T prediction of $\varepsilon_c$ to the behavior of the NL system in Figure \ref{fig:NLvsS3T_stabilityboundary}. 
Because the NL system fluctuates stochastically, a precise NL stability boundary does not exist so we instead identify the onset of VSHF formation with an abrupt increase of zmf, previously shown in Figure \ref{fig:modelcomparison} (b). 
We carried out an NL simulation for each $(\varepsilon,N_0^2)$ pair on the grid defined by $N_0^2=10^i$, $\varepsilon = 10^j$, with $i=1,1.5,\ldots,4$, and $j=0,0.25,\ldots3$. 
The simulations were spun up for 450 time units at low resolution before simulating 20 time units at $N=512$ resolution over which the mean zmf value was calculated. 
The thick dark curve in Figure \ref{fig:NLvsS3T_stabilityboundary} shows $\varepsilon_c$ as predicted\footnote{To calculate the value of $\varepsilon_c$ appropriate for comparison to simulations we apply (\ref{eq:varepsc_minimization}) with $m$ restricted to only the values permitted by the doubly periodic domain.} by S3T and the shaded contours show the time average zmf values from the NL simulations. 

For intermediate and strong stratification, $\varepsilon_c$ provides a good prediction of the onset of VSHF formation in NL. 
In particular, the NL system verifies the S3T predictions that VSHF formation occurs most readily at intermediate stratification and that the excitation strength at which VSHFs form grows like $N_0^2$ as the stratification becomes strong. 
 
Another feature visible in Figure \ref{fig:NLvsS3T_stabilityboundary}, and also in Figure \ref{fig:modelcomparison} (b) which is a `slice' through Figure \ref{fig:NLvsS3T_stabilityboundary} at $N_0^2=10^3$, is that the NL zmf reaches a maximum and subsequently decreases as $\varepsilon$ is increased. 
This finite-amplitude effect is outside the scope of the present work, which analyzes the VSHF-forming instability from a linear perspective. 
However, inspection of the individual simulations suggests that the eventual decrease of zmf may be due to the VSHF maintaining a relatively large vertical wavenumber as $\varepsilon$ is increased. 
This behavior is in contrast to the usual observation in $\beta$-plane turbulence that jets transition to lower wavenumber as the excitation strength is increased \citep{Farrell:2007fq}. 
Our previous analysis of VSHF formation with matrix S3T is consistent with this interpretation, and also suggests that additional turbulent equilibria consisting of lower-wavenumber, more energetic VSHFs may be simultaneously stable with the equilibria shown in Figure \ref{fig:NLvsS3T_stabilityboundary} \citep{Fitzgerald:2016ux}.


Under weak stratification ($N_0^2\lesssim 10$), the VSHF does not obtain significant energy in our NL simulations, consistent with previous studies \citep{Smith:2001uo,Kumar:2017ie}. 
An energetic VSHF creates strong shear, which for weak stratification is associated with hydrodynamic instability via the Miles-Howard (MH) criterion. 
Although the MH criterion is formally valid only for a static parallel flow in the absence of excitation and dissipation, it provides a useful guide for intuition and suggests that maintaining strong VSHFs at weak stratification is unlikely, as instability of the emergent jet is likely to prevent its obtaining significant amplitude. 

\subsection{MCE}

The lighter curves in Figure \ref{fig:StabilityBoundaryAsymptotics} (a) show the stability boundaries in the MCE cases. 
As in the case of IRE, the stability boundaries reflect the properties of $s_U$ discussed in Section \ref{sec:dispersion}\ref{sec:dispersionMCE}. 
Striking differences are evident between the MCE$_2$ and MCE$_{1/2}$ cases. 
As $N_0^2\to 0$, $s_U$ remains positive for MCE$_2$ and remains negative for MCE$_{1/2}$ (see Figure \ref{fig:s_vs_n} (b,c)). 
As a result, for small $N_0^2$, $\varepsilon_c$ tends to a constant value for MCE$_2$ while $\varepsilon_c$ does not exist for MCE$_{1/2}$ as the VSHF-forming instability does not occur for any excitation strength until the stratification becomes strong. 
The stability boundaries also reflect the surprising result, previously shown in Figure \ref{fig:LargeNsqAsymp}, that strongly stratified homogeneous turbulence maintained with MCE$_{1/2}$ is more unstable to the VSHF-forming instability than that maintained with MCE$_2$ or IRE, even though the  instability does not occur at all for MCE$_{1/2}$ when the stratification is weak. 

The emergent VSHF wavenumber, $m^{\star}$, also differs markedly between the MCE cases (Figure \ref{fig:StabilityBoundaryAsymptotics} (b), lighter curves). 
For small $N_0^2$, MCE$_2$ has $m^{\star}\approx 0.6$ so that the vertical scale of the emergent VSHF is comparable to the horizontal scale of the excitation, $1/k_0$.  
When the VSHF-forming instability first occurs for MCE$_{1/2}$ near $N_0^2=10^2$, the VSHF instead has $m^{\star}\approx 1.6$, so that the vertical scale of the VSHF is smaller than $1/k_0$ and is associated with the correlation length of the excitation in the vertical, $\ell_c$.  
For large $N_0^2$ the MCE and IRE cases all behave similarly, with $m^{\star}\to0$ in all cases, consistent with the results shown in Figure \ref{fig:LargeNsqAsymp}.   

In the remaining sections, we revisit these observations and analyze their dynamical origins from the perspective of wave-mean flow interaction. 

\section{Feedback factors} \label{sec:feedback}

Because the properties of the VSHF-forming instability can depend on the excitation structure, it is useful to analyze the instability from a perspective independent of the particular excitation. 
The feedback factor, first developed by \citet{Bakas:2015iy} in the context of the zonostrophic instability, provides a tool for analyzing VSHF formation in this way. 
In the feedback factor approach, the strength and sign of the feedback resulting from the interaction between the VSHF and each wavenumber component of the turbulence spectrum is analyzed independently. 
Each spectral component either supports or opposes VSHF development, and the total wave-mean flow feedback for a particular excitation is given by the sum of the feedbacks arising from each component. 
This perspective facilitates understanding the properties of the VSHF-forming instability demonstrated in Section \ref{sec:dispersion}.

We focus on the case of a stationary, wavenumber $m$ VSHF of perturbative amplitude at its stability boundary, so that $U=\delta U$, $\partial_t=s_U=0$, and $\varepsilon=\varepsilon_c(m)$. 
From (\ref{eq:NL3}), the Reynolds stress and VSHF structures are related as $r_m \delta U = -\partial_z \langle u'w'\rangle$. 
Combining this with (\ref{eq:stabilityboundary}) we obtain
\begin{equation}
-\frac{\partial}{\partial z} \langle u'w' \rangle = \delta U \varepsilon_c(m) \iint \text{d}p\text{d}q\left(2r\mathcal{F}_U\Big|_{s_U=0}\right) \tilde{E}. \label{eq:reynoldsstressFU}
\end{equation}
The induced Reynolds stress in (\ref{eq:reynoldsstressFU}) scales linearly with $\delta U$, which follows from linearization, and also with $\varepsilon$, which follows from the quasilinearity of the dynamics underlying S3T. 
The remaining factor in (\ref{eq:reynoldsstressFU}), which determines the sign of the induced stress, is the integral over the excitation spectrum, $\tilde{E}$, weighted by the feedback factor for each spectral component, $2r\mathcal{F}_U$, evaluated at $s_U=0$. 
We hereafter refer to $\mathcal{F}_U$ as the feedback factor for simplicity, as the $2r$ factor scales the amplitude but does not modify the structure, and suppress the notation indicating that $\mathcal{F}_U$ is evaluated at $s_U=0$. 

The feedback factor depends on the four arguments $(p,q,N_0^2,m)$ and characterizes the wave-mean flow feedback occurring between waves excited with wavenumber $(p,q)$ and a weak VSHF with wavenumber $m$. 
If the net feedback is positive when integrated over the excited spectrum in (\ref{eq:reynoldsstressFU}), the induced Reynolds stress divergence is proportional to $\delta U$ and reinforces the VSHF so that VSHFs at wavenumber $m$ grow for sufficiently large $\varepsilon$. 
If the net feedback is negative, the Reynolds stress divergence opposes the VSHF and VSHFs at wavenumber $m$ decay faster than $r_m$. 
The feedback factor thus underlies VSHF formation and understanding the structure of $\mathcal{F}_U$ is central for understanding the VSHF-forming instability. 

\begin{figure}[t]
\centerline{\includegraphics[scale=1]{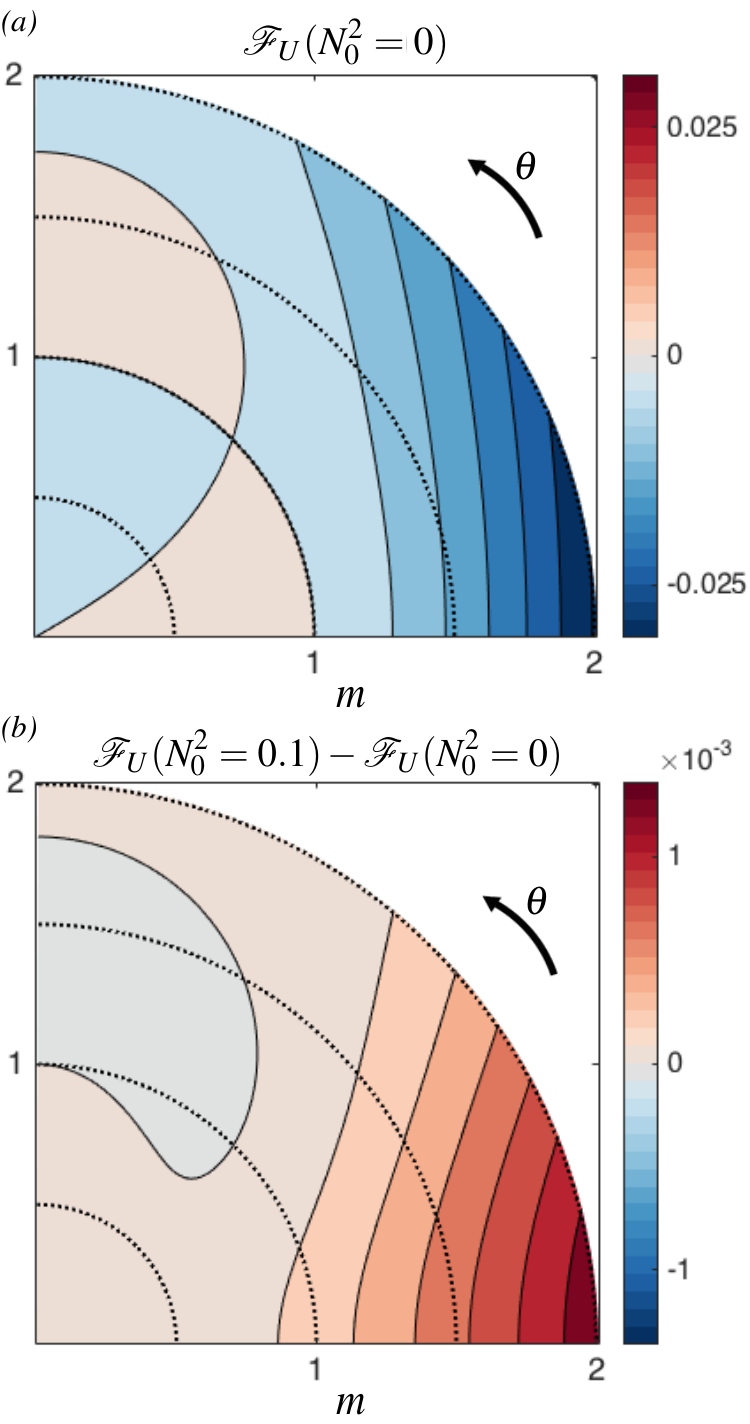}}
\caption{Wave-mean flow feedback factor, $\mathcal{F}_U$, for the VSHF-forming instability for the cases of zero and weak stratification, shown in polar coordinates appropriate for analyzing IRE. The radial coordinate indicates the VSHF wavenumber, $m$, and the polar angle indicates the angle of the excited wave according with the convention $(p,q)=(\cos\theta,\sin\theta)$. Panel (a) shows $\mathcal{F}_U$ for $N_0^2=0$ and panel (b) shows how $\mathcal{F}_U$ is modified by the introduction of weak stratification. The parameters used are $r_m=0.1$ and $\nu=0$.} \label{fig:FeedbackFactorsPolar} 
\end{figure}

The complete structure of $\mathcal{F}_U$ cannot be visualized at once due to its many arguments. 
However, when the excitation structure is particularly simple, such as IRE and MCE which excite 1D subspaces of the available 2D spectrum, the relevant $\mathcal{F}_U$ structure can be visualized easily. 
Figure \ref{fig:FeedbackFactorsPolar} shows the $\mathcal{F}_U$ structure relevant to IRE in polar coordinates, with the radial coordinate indicating $m$ and the polar angle indicating the angle of the excited wave, $\theta$, where $(p,q)=(\cos\theta,\sin\theta)$. 
As IRE is doubly mirror-symmetric, with $\tilde{E}(p,q)=\tilde{E}(-p,q)=\tilde{E}(p,-q)=\tilde{E}(-p,-q)$, we sum the contributions to $\mathcal{F}_U$ from each of these related Fourier components and plot the resulting $\mathcal{F}_U$ values over the first quadrant, $0\le\theta\le \pi/2$. 
Red regions in Figure \ref{fig:FeedbackFactorsPolar} correspond to $\mathcal{F}_U>0$ and produce Reynolds stresses that reinforce the VSHF, while blue regions correspond to the opposite case, $\mathcal{F}_U<0$. 
The net wave-mean flow interaction between the IRE spectrum and a VSHF with wavenumber $m_0$ is determined by comparing the size and strength of the $\mathcal{F}_U>0$ and $\mathcal{F}_U<0$ regions over the contour $m=m_0$. 


We showed in Figure \ref{fig:SmallNsqAsymp} that unstratified IRE turbulence has no net influence on VSHFs with $0<m<1$, so that $s_U=-r_m$, and that turbulence opposes VSHFs with $m>1$, so that $s_U<-r_m$. 
Figure \ref{fig:FeedbackFactorsPolar} (a) provides an explanation for this behavior. 
VSHFs with $0<m<1$ are supported by waves with small $\theta$ and opposed by waves with larger $\theta$. 
These competing effects cancel exactly, resulting in zero net feedback. 
For $m>1$ the range of $\theta$ for which $\mathcal{F}_U<0$ widens, resulting in negative net feedback. 

We also showed in Figure \ref{fig:SmallNsqAsymp} that for weak but nonzero stratification, $s_U$ is enhanced for $0<m<1$ relative to the $N_0^2=0$ case, with the strongest enhancement at $m=1$.  
The changes in $\mathcal{F}_U$ as $N_0^2$ is increased from zero, shown in Figure \ref{fig:FeedbackFactorsPolar} (b), explain this observation. 
The symmetry between the $\mathcal{F}_U>0$ and $\mathcal{F}_U<0$ regions for $0<m<1$ is broken when $N_0^2>0$, favoring $\mathcal{F}_U>0$ for most $(m,\theta)$ pairs but especially near $m\approx1$ for small $\theta$.  
S3T thus predicts that an $m=1$ VSHF emerges in weakly stratified IRE turbulence near the stability boundary. 

\begin{figure}[t]
\centerline{\includegraphics[scale=1]{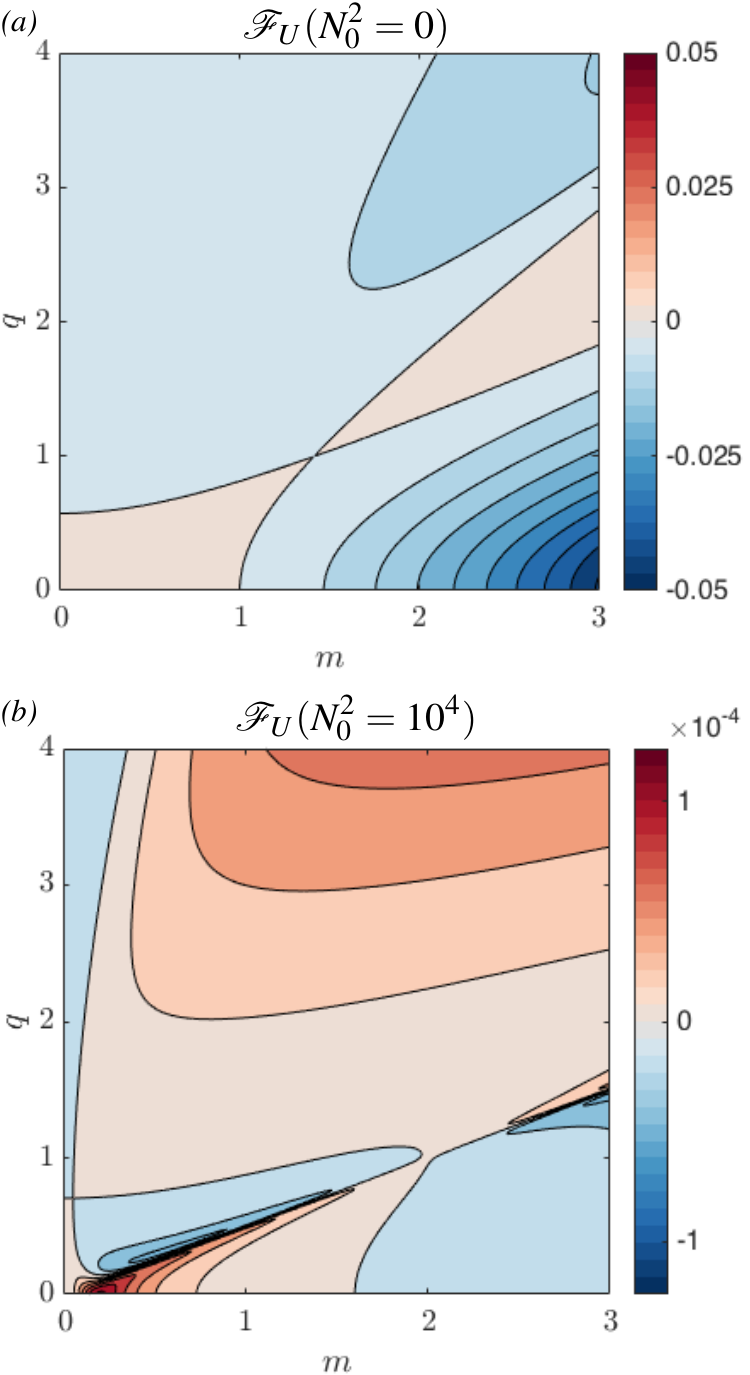}}
\caption{Feedback factor, $\mathcal{F}_U$, for the VSHF-forming instability in Cartesian coordinates appropriate for analyzing MCE. The horizontal axis indicates the VSHF wavenumber, $m$, and the vertical axis indicates the vertical wavenumber of the excited wave, $q$. Panel (a) shows $\mathcal{F}_U$ in the unstratified case ($N_0^2=0)$ and panel (b) shows $\mathcal{F}_U$ under strong stratification ($N_0^2=10^4$). The parameters used are $r_m=0.1$ and $\nu=0$.} \label{fig:Monok_FU} 
\end{figure}

\begin{figure}[t]
\centerline{\includegraphics[scale=1]{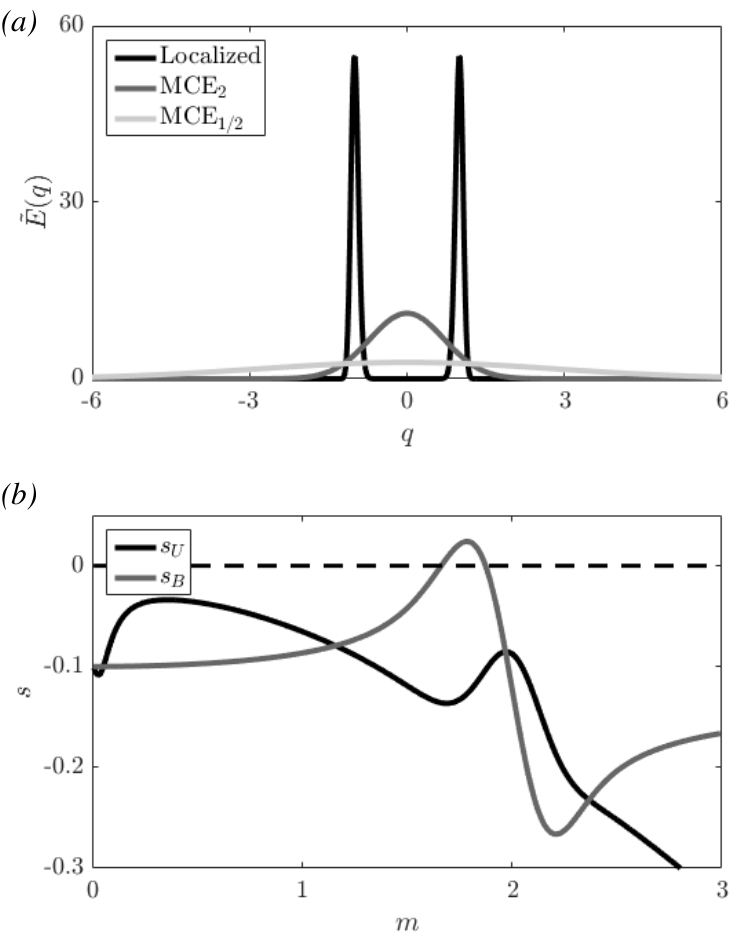}}
\caption{Example demonstrating that the buoyancy layering instability can occur for appropriately chosen excitation. Panel (a) shows the energy injection spectrum for the localized excitation chosen to induce buoyancy layering alongside the MCE$_2$ and MCE$_{1/2}$ spectra which are provided for reference. Panel (b) shows the growth rates of the VSHF-forming and buoyancy layering instabilities as functions of $m$ for the localized excitation with parameters $\varepsilon=400$ and $N_0^2=10^4$. The most unstable structure corresponds to buoyancy layers with $m\approx1.8$, which emerge from homogeneous turbulence as an S3T instability. The parameters used are $r_m=0.1$ and $\nu=0$.} \label{fig:Layering} 
\end{figure}

Figure \ref{fig:StabilityBoundaryAsymptotics} showed that the VSHF-forming instability occurs for MCE$_{2}$ when $N_0^2=0$ but does not occur for MCE$_{1/2}$ until the stratification is strong. 
Figure \ref{fig:Monok_FU} shows the $\mathcal{F}_U$ structure that underlies this behavior, now using Cartesian coordinates appropriate for the MCE in which the horizontal axis indicates the VSHF wavenumber, $m$, and the vertical axis indicates the vertical wavenumber of the excited wave, $q$. 
As in the case of IRE, we sum the contributions to $\mathcal{F}_U$ from Fourier components that are related by the double mirror symmetry of MCE and plot the result over $q>0$. 
For reference, Figure \ref{fig:Layering} (a) shows the 1D energy injection spectra, $\tilde{E}(q)$, for MCE$_2$ and MCE$_{1/2}$. 
MCE$_2$ primarily excites $q\lesssim1$ while MCE$_{1/2}$ injects significant energy over $q\lesssim4$. 
The $\mathcal{F}_U(N_0^2=0)$ structure in Figure \ref{fig:Monok_FU} (a) shows that the $q<1$ region excited by MCE$_2$ predominantly has $\mathcal{F}_U>0$ for $m<1$, explaining the occurrence of the VSHF-forming instability for unstratified MCE$_2$ with $m^{\star}\approx0.6$. 
Although MCE$_{1/2}$ excites the same $q<1$ waves, it also excites a broad band of $q>1$ waves with $\mathcal{F}_U<0$, resulting in $s_U<0$ for $N_0^2=0$. 


In Section \ref{sec:stabilityboundary} we discussed the surprising result that, under strong stratification, the VSHF-forming instability occurs at lower $\varepsilon$ for MCE$_{1/2}$ than for MCE$_2$ (Figure \ref{fig:StabilityBoundaryAsymptotics} (a)). 
The structure of $\mathcal{F}_U$ for large $N_0^2$, shown in Figure \ref{fig:Monok_FU} (b), explains this phenomenon. 
For large $N_0^2$, a broad band of $\mathcal{F}_U>0$ waves exists for $q>1$. 
This band is excited by MCE$_{1/2}$, leading to relatively large VSHF growth rates for large $N_0^2$. 
The $q<1$ band excited by MCE$_2$ exhibits a dipole structure in which waves that strongly reinforce the VSHF compete with others that strongly oppose it, weakening the net feedback and the instability growth rate for MCE$_2$.

\begin{figure}[t]
\centerline{\includegraphics[scale=1]{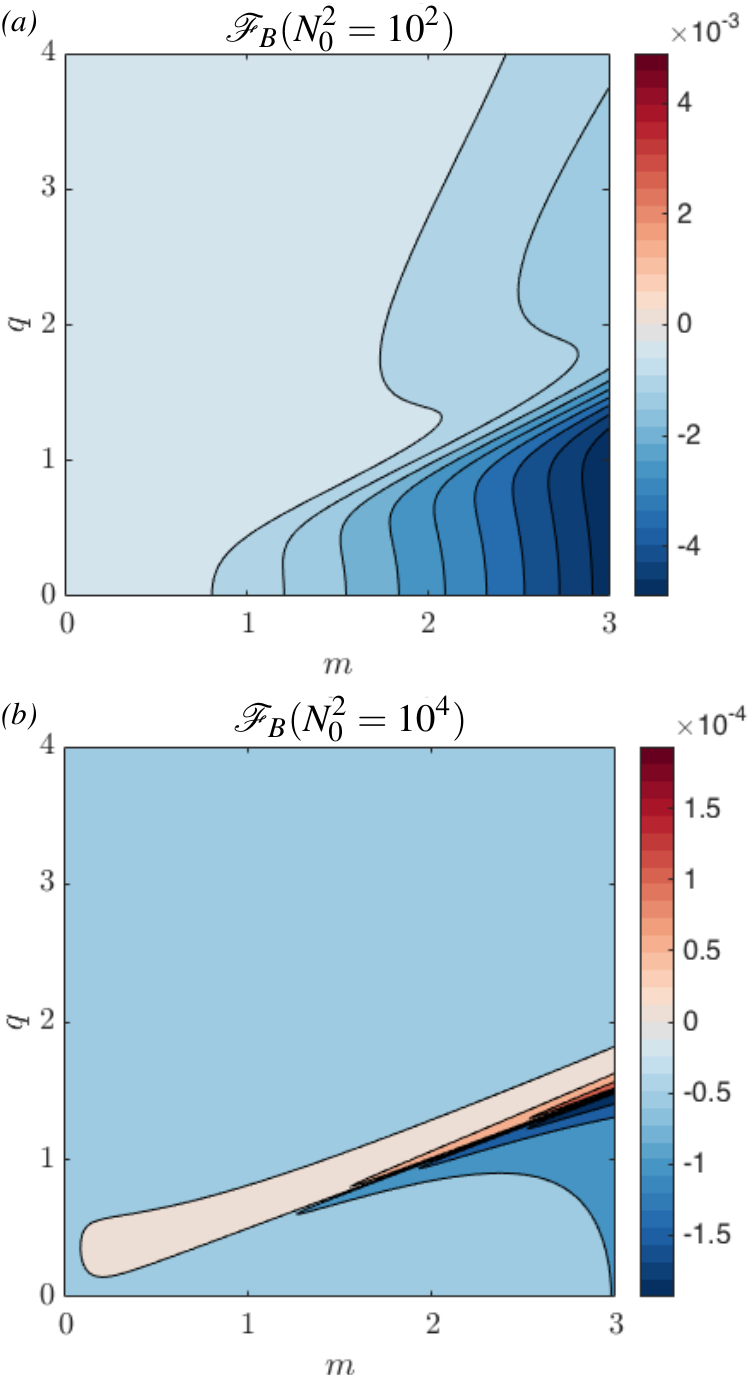}}
\caption{Feedback factor, $\mathcal{F}_B$, for the buoyancy layering instability in Cartesian coordinates as in Figure \ref{fig:Monok_FU}. Panel (a) shows $\mathcal{F}_B$ under intermediate stratification ($N_0^2=10^2)$ and panel (b) shows $\mathcal{F}_B$ under strong stratification ($N_0^2=10^4)$. This figure shows that the feedback is usually negative, consistent with the observation that the buoyancy layering instability does not occur for IRE or MCE. However, panel (b) shows that a band of positive feedback exists in the strongly stratified case. The parameters used are $r_m=0.1$ and $\nu=0$.} 
\label{fig:Monok_FB} 
\end{figure}

Feedback factor analysis can also be applied to the buoyancy layering instability. 
Following the approach for the VSHF-forming instability, the feedback factor $\mathcal{F}_B(p,q,N_0^2,m)$ characterizes the feedback between waves excited at wavenumber $(p,q)$ and weak buoyancy layers with wavenumber $m$. 
The structure of $\mathcal{F}_B$ is shown in Figure \ref{fig:Monok_FB}, revealing that the feedback is usually negative, with a narrow band of positive interactions emerging at strong stratification. 
As a result, the buoyancy layering instability fails to occur for either IRE or MCE, which do not preferentially excite the $\mathcal{F}_B>0$ band available at strong stratification. 


The existence of an $\mathcal{F}_B>0$ band in Figure \ref{fig:Monok_FB} (b) raises the intriguing possibility that preferential excitation of this band might produce turbulence in which the buoyancy layering instability occurs. 
In Figure \ref{fig:Layering} we demonstrate that this instability indeed occurs with a faster growth rate than the VSHF-forming instability when the excitation is carefully chosen. 
Panel (a) shows the chosen excitation, which is highly localized near $q=1$. 
Waves excited with $q\approx1$ engage in a strong positive feedback with $m\approx2$ buoyancy layers.  
Although these waves also engage in a positive feedback with $m\lesssim1$ VSHFs, the net buoyancy layering feedback is stronger than the net VSHF-forming feedback. 
Panel (b) shows $s_U$ and $s_B$ as functions of $m$ for the localized excitation. 
As anticipated from feedback factor analysis, the fastest growing coherent structures are buoyancy layers with $m\approx 1.8$, and $s_U<0$ for all $m$. 
It is thus possible for buoyancy layers to spontaneously form in homogeneous turbulence via quasilinear interactions between the emergent layers and the wave field. 
However, because the turbulent spectrum required for this scenario is highly contrived, this mechanism is unlikely to be found in nature or in numerical simulations using more natural excitation.


\section{Processes Contributing to Wave-Mean Flow Feedback} \label{sec:processes}

The quasilinear wave-mean flow feedback mechanism characterized by $\mathcal{F}_U$ operates via several physical processes. 
One such process, sometimes referred to as the Orr mechanism, is shear straining of vorticity perturbations by the VSHF to produce upgradient momentum fluxes. 
We now briefly analyze the contributions of individual processes to the VSHF-forming instability. 
This analysis reveals that different processes can act as the dominant driver of VSHF formation for different choices of excitation. 

The Reynolds stresses that reinforce the VSHF during its exponential growth phase are associated, through (\ref{eq:eddymomentumflux}), with perturbations to the vorticity covariance, $\delta Z$. 
Three quasilinear processes, represented by terms in (\ref{eq:S3T_lin_phys_1_MAIN}), produce structure in $\delta Z$ that yields Reynolds stresses. 
The first process, represented by the term $(\delta U_1 -\delta U_2)\partial_x Z_H$, is the previously described Orr mechanism. 
The second process, represented by the term $-(\delta U_1'' -\delta U_2'')\Delta \partial_x \Psi_H$, is the advection of the VSHF vorticity by the perturbations. 
The third process, represented by the term $-\partial_x(\delta\Gamma^b-\delta \Gamma^{\zeta})$, is the production of vorticity perturbations by buoyancy perturbations. 
The third process is the most complex as it subsumes a variety of processes involving vorticity-buoyancy coupling such as gravity wave dynamics. 
The feedback factor, $\mathcal{F}_U$, can be decomposed into contributions from each of these processes as \begin{equation}
\mathcal{F}_U = \mathcal{F}_U^{Orr}+\mathcal{F}_U^{cu}+\mathcal{F}_U^{wave}, \label{eq:FU_decomp}
\end{equation}
where the superscripts identify the component feedbacks resulting from the Orr mechanism (Orr), from advection of VSHF vorticity by perturbations (cu, for curvature), and from vorticity-buoyancy coupling including gravity wave dynamics (wave). 
Mathematical details are provided in Appendix D. 
We note that only $\mathcal{F}_U^{wave}$ depends on the stratification, $N_0^2$, as the Orr and curvature feedbacks do not involve vorticity-buoyancy coupling. 

\begin{figure}[t]
\centerline{\includegraphics[scale=1]{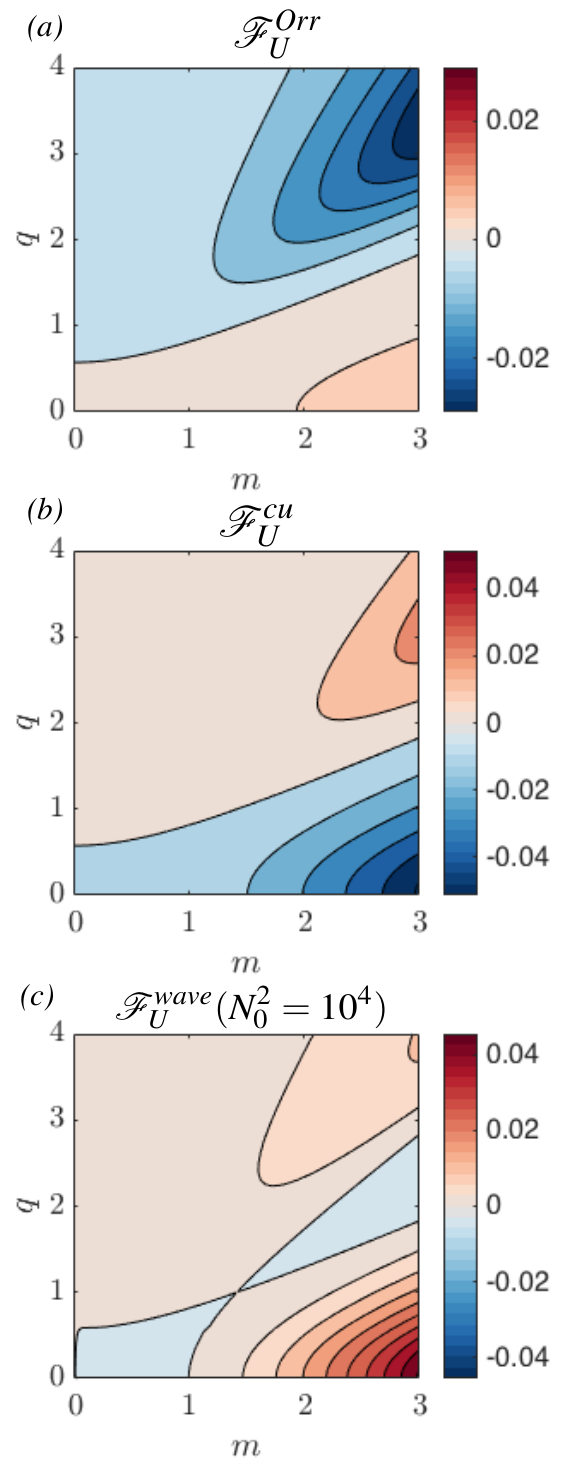}}
\caption{Decomposition of the feedback factor, $\mathcal{F}_U$, for the VSHF-forming instability into its contributions from the Orr (a), curvature (b), and wave (c) feedbacks as in (\ref{eq:FU_decomp}). Axes are as in Figure \ref{fig:Monok_FU}. The Orr and curvature feedbacks are independent of $N_0^2$ and the wave feedback is shown under strong stratification ($N_0^2=10^4$). This decomposition shows that the $m\approx0.5$ VSHF emerging for MCE$_{2}$, which primarily excites $q\lesssim1$, is primarily driven by the Orr mechanism, while the $m\approx1.5$ VSHF emerging for MCE$_{1/2}$, which excites $q\lesssim4$, is primarily driven by vorticity-buoyancy coupling. The parameters used are $r_m=0.1$ and $\nu=0$.} \label{fig:Mechanism_Figure} 
\end{figure}

To illustrate this technique we apply (\ref{eq:FU_decomp}) to VSHF formation in the case of MCE. 
Figure \ref{fig:Mechanism_Figure} shows the contribution of each process to $\mathcal{F}_U$ for $N_0^2=10^4$. 
In Section \ref{sec:stabilityboundary} we showed that MCE$_2$, which primarily excites waves with $q<1$, forms a VSHF with $m\approx0.5$ for these parameter values, while MCE$_{1/2}$, which excites waves with $q\lesssim4$, forms a VSHF with $m\approx1.5$. 
Inspection of the region $m \approx 0.5$, $0<q<1$ in Figure \ref{fig:Mechanism_Figure} indicates that VSHF formation for MCE$_2$ is driven by the Orr mechanism, and that the curvature and wave feedbacks oppose VSHF formation. 
In contrast, the region $m \approx 1.5$, $q\lesssim4$ in Figure \ref{fig:Mechanism_Figure} indicates that VSHF formation for MCE$_{1/2}$ is driven by the wave feedback. 
The net feedbacks from Orr and curvature dynamics result from a competition between negative and positive feedbacks from different parts of the spectrum, and detailed integration reveals that both processes oppose VSHF formation. 
The quasilinear feedback mechanism thus produces VSHF formation for both MCE$_2$ and MCE$_{1/2}$ but exploits distinct physical processes in each case. 


\section{Discussion} \label{sec:discussion}

In this work we applied S3T to analyze VSHF formation in 2D stratified turbulence. 
We focused on the initial VSHF emergence in homogeneous turbulence maintained by stochastic excitation. 
VSHF emergence occurs through an S3T instability of homogeneous turbulence which we refer to as the VSHF-forming instability.  
Some properties of the VSHF-forming instability, such as the shape of the stability boundary, the scale of the emergent VSHF, and the detailed physical mechanism of the instability depend on the structure of the stochastic excitation.
We explained these properties in terms of the statistical wave-mean flow feedback mechanism which drives VSHF formation and the basic physical processes that underlie the feedback. 
Our analysis complements recent work in which we applied S3T to analyze VSHFs at finite amplitude \citep{Fitzgerald:2016ux}. 

Our analysis extended to the VSHF-forming instability several S3T concepts and techniques developed in the context of the zonostrophic instability in $\beta$-plane turbulence. 
In particular, a primary contribution of this work was to extend to the VSHF-forming instability the differential linearized formulation of S3T due to \citet{Srinivasan:2012im}. 
The differential approach to S3T is invaluable for understanding the initial emergence of coherent structure in turbulence because it allows the parameter dependence and asymptotic behavior to be analyzed using closed-form expressions. 
We emphasize, however that this approach is formally equivalent to the conventional matrix implementation of S3T. 
\citet{StOnge:2017tu} recently extended the differential S3T approach to the turbulence of stochastically excited interchange modes in plasmas. 
This turbulence is equivalent to stochastically excited Rayleigh-B\'{e}nard convection for subcritical Rayleigh number, which is a weakly unstably stratified turbulence closely related to the stably stratified turbulence that we analyze. 

The VSHF-forming instability is revealed by our analysis to be similar to the zonostrophic instability in several respects. 
Comparison of the VSHF-forming and zonostrophic instabilities reveals that the role played by the stratification, $N_0^2$, in the VSHF-forming instability is instead played by the planetary vorticity gradient, $\beta$, in zonostrophic instability. 
For example, for IRE turbulence the zonostrophic instability growth rate decays like $1/\beta^2$ as $\beta\to\infty$ and increases from zero like $\beta^2$ for small $\beta$ in the absence of explicit jet damping. 
\citet{Bakas:2015iy} also showed that the properties of the zonostrophic instability depend on the structure of the stochastic excitation. 
In particular, structures that primarily excite Rossby waves with nearly horizontal wavevectors produce positive zonostrophic instability growth rates as $\beta\to0$, whereas the zonostrophic instability does not occur at all for weak $\beta$ for structures that primarily excite waves with nearly vertical wavevectors. 
These properties mirror the properties of the VSHF-forming instability analyzed in this paper. 
From the S3T perspective, zonal jet emergence in geostrophic turbulence and VSHF emergence in non-rotating stratified turbulence can be usefully conceptualized as two instances of the more generic phenomenon of S3T instability of homogeneous turbulence. 


\acknowledgments
The authors thank N.\ C.\ Constantinou for helpful discussions, J.\ R.\ Taylor for providing the DIABLO code, and two anonymous reviewers for useful comments on the manuscript. J.\ G.\ F.\ was partially supported by a doctoral fellowship from the National Sciences and Engineering Research Council of Canada. B.\ F.\ F.\ was partially supported by the U.S. National Science Foundation under Grant Nos. NSF AGS-1246929 and NSF AGS-1640989.

\appendix[A] \label{sec:AppendixDispersion2}
\appendixtitle{Dispersion Relations} 

In this Appendix we provide additional details regarding the derivation of equations (\ref{eq:dispersion_U}) and (\ref{eq:dispersion_B}) for the growth rates of the VSHF-forming and buoyancy layering instabilities. 

Linearizing (\ref{eq:S3T1})-(\ref{eq:S3T3}) about the fixed point corresponding to homogeneous turbulence given by (\ref{eq:homoFP}), we obtain 
\begin{multline}
\partial_t \delta Z + (\delta U _1 -\delta U _2 )\partial_x Z_H - (\delta U_1''-\delta U_2 '' )\Delta \partial_x \Psi_H \\
=-2[r-\nu(\Delta+\frac14\partial^2_{\bar{z}\bar{z}})]\delta Z+\partial_x \delta \Gamma^{\text{diff}},  \label{eq:S3T_lin_phys_1} 
\end{multline}
\begin{multline}
\partial_t \delta T + (\delta U_1 - \delta U_2)\partial_x T_H - N_0^2 \partial_x \delta S^{\text{diff}}  = \\ 
-2[r-\nu(\Delta+\frac14\partial^2_{\bar{z}\bar{z}})]\delta T,  \label{eq:S3T_lin_phys_2}
\end{multline}
\begin{multline} 
\partial_t \delta \Gamma^{\text{sum}} - 2N_0^2\partial^3_{xz\bar{z}}\delta \Psi + (\delta B_1'-\delta B_2')\Delta \partial_x \Psi_H =  \\
-2[r-\nu(\Delta+\frac14\partial^2_{\bar{z}\bar{z}})]\delta \Gamma^{\text{sum}},  \label{eq:S3T_lin_phys_3} 
\end{multline}
\begin{multline}
\partial_t \Gamma^{\text{diff}} +2N_0^2(\Delta + \frac14 \partial^2_{\bar{z}\bar{z}})\partial_x \delta \Psi + (\delta B_1' + \delta B_2')\Delta \partial_x \Psi_H \\
= -2[r-\nu(\Delta+\frac14\partial^2_{\bar{z}\bar{z}})] \delta \Gamma^{\text{diff}} -2\partial_x \delta T,  \label{eq:S3T_lin_phys_4} 
\end{multline}
\begin{align}
\partial_t \delta U &= (-r_m+\nu\partial^2_{\bar{z}\bar{z}}) \delta U - \partial^3_{xz\bar{z}} \delta \Psi \Big|_{x=z=0},  \label{eq:S3T_lin_phys_5}  \\
\partial_t \delta B &= (-r_m+\nu\partial^2_{\bar{z}\bar{z}})\delta B + \frac12 \partial^2_{x\bar{z}} \delta S^{\text{diff}} \Big|_{x=z=0}.  \label{eq:S3T_lin_phys_6} 
\end{align}
In these equations we have included the viscosity terms which were omitted for clarity in the main text and have, for convenience, expressed the dynamics in terms of the quantities 
\begin{align}
\Gamma^{\text{sum}}\equiv \Gamma^b+\Gamma^{\zeta}, && \Gamma^{\text{diff}}\equiv \Gamma^b-\Gamma^{\zeta}, \\
S^{\text{sum}}\equiv S^b + S^{\zeta}, && S^{\text{diff}}\equiv S^b - S^{\zeta},
\end{align}
which are related through the expressions
\begin{align}
\Gamma^{\text{sum}} &= (\Delta+\frac14 \partial^2_{\bar{z}\bar{z}})S^{\text{sum}}-\partial^2_{z\bar{z}}S^{\text{diff}}, \\
\Gamma^{\text{diff}} &= (\Delta+\frac14 \partial^2_{\bar{z}\bar{z}})S^{\text{diff}}-\partial^2_{z\bar{z}}S^{\text{sum}}.
\end{align}

We analyze the linearized system (\ref{eq:S3T_lin_phys_1})-(\ref{eq:S3T_lin_phys_6}) in Fourier space, using the ansatz (\ref{eq:FourierAnsatz1})-(\ref{eq:FourierAnsatz3}) and further writing $\hat{C}_{m,s}(x,z)$, the homogeneous structure of the perturbation to the covariance function, using its Fourier transform
\begin{equation}
\hat{C}_{m,s}(x,z)=\iint \frac{\mbox{d}p \mbox{d}q}{(2\pi)^2} e^{i(px+qz)}\tilde{C}(p,q)_{m,s}.
\end{equation}
Using the relations 
\begin{align}
\delta U_1 - \delta U_2 &= 2i\sin(mz/2)e^{im\bar{z}}e^{st} \hat{U}_{m,s}, \\
\delta U_1'' - \delta U_2'' &= -2im^2\sin(mz/2)e^{im\bar{z}}e^{st} \hat{U}_{m,s}, \\
\delta B_1' - \delta B_2' &= -2m\sin(mz/2)e^{im\bar{z}}e^{st} \hat{B}_{m,s}, \\
 \delta B_1' + \delta B_2' &= 2im\cos(mz/2)e^{im\bar{z}}e^{st} \hat{B}_{m,s},
\end{align}
we obtain the linearized dynamics in Fourier space as 
\begin{flalign}
0&=s' \tilde{Z}-ip\hat{U}(\tilde{\Phi}_H^+-\tilde{\Phi}^-_H) - ip\tilde{\Gamma}^{\text{diff}} \label{eq:Ztildefinal} && \\
0&=s' \tilde{T}-ip\hat{U}(\tilde{T}_H^+-\tilde{T}^-_H)-ipN_0^2\tilde{S}^{\text{diff}}  \label{eq:Ttildefinal} && \\
0&=s' \tilde{\Gamma}^{\text{sum}} +2impqN_0^2 \tilde{\Psi} +mp\hat{B}(\tilde{X}_H^+-\tilde{X}_H^-)  \label{eq:Ktildeplusfinal} && \\
0&=s' \tilde{\Gamma}^{\text{diff}} - 2ipN_0^2(h^2+\frac{m^2}{4})\tilde{\Psi}-mp\hat{B}(\tilde{X}_H^-+\tilde{X}_H^+)+2ip\tilde{T} &&\label{eq:Ktildeminusfinal} \\
\bar{s} \hat{U} &= im\iint \frac{\mbox{d}p \mbox{d}q}{(2\pi)^2} pq \tilde{\Psi} \label{eq:Ufinal} && \\
\bar{s} \hat{B} &= -\frac{m}{2}\iint \frac{\mbox{d}p\mbox{d}q}{(2\pi)^2} p \tilde{S}^{\text{diff}} && \label{eq:Bfinal}
\end{flalign}
where we have suppressed the $m,s$ subscripts, defined the quantities $s'=s+2(r+\nu(h^2+m^2/4))$, $\bar{s}=s+r_m+\nu m^2$, $\Phi_H = (\Delta^2 + m^2 \Delta) \Psi_H$, and $X_H = \Delta \Psi_H$, and introduced the notation $\tilde{f}^{\pm}=\tilde{f}(p,q\pm\frac{m}{2})$. 

We now manipulate (\ref{eq:Ztildefinal})-(\ref{eq:Bfinal}) to obtain the dispersion relations. 
The dispersion relations for the VSHF-forming and buoyancy layering instabilities can be obtained separately because the eigenproblem defined by (\ref{eq:Ztildefinal})-(\ref{eq:Bfinal}) factors into two decoupled eigenproblems, one for VSHFs and one for buoyancy layers, under the assumptions that the excitation satisfies the equal energy and non-correlation condition (\ref{eq:ExcitationAssumption}) and the reflection symmetry $\Xi(p,q)=\Xi(-p,q)$. 
This factorization property can be verified after obtaining the dispersion relations by confirming that the perturbations to the covariance matrix associated with VSHF formation produce no eddy buoyancy flux divergences, and vice versa for those associated with buoyancy layering. 
 
The dispersion relation for the VSHF-forming instability is obtained by setting $\hat{B}=0$ in (\ref{eq:Ztildefinal})-(\ref{eq:Bfinal}) so that the horizontal mean structure corresponds to a VSHF with no mean buoyancy perturbation.
Equations (\ref{eq:Ztildefinal})-(\ref{eq:Ktildeminusfinal}) can then be solved for $\tilde{\Psi}$ to obtain 
\begin{equation}
\tilde{\Psi} = ips'_U \hat{U} \frac{ ( \tilde{\Phi}_H^+-\tilde{\Phi}_H^- ) F_0+(\tilde{T}_H^{+}-\tilde{T}_H^{-})(2p^2h_{-}^2h_{+}^2)}{F_0^2-4p^4N_0^4h_{-}^2h_+^2}, \label{eq:PsiSolution}
\end{equation}
where $F_0=  s'^2_Uh_{-}^2h_+^2+2p^2N_0^2(h^2+\frac{m^2}{4})$ and $h^2_{\pm}=p^2+(q\pm \frac12 m)^2$. 
The assumption that $\hat{B}=0$ can be shown to be consistent by similarly solving (\ref{eq:Ztildefinal})-(\ref{eq:Ktildeminusfinal}) for $\tilde{S}^{\text{diff}}$ (not shown) and substituting the result into (\ref{eq:Bfinal}). 
Inspection of the right-hand side of the resulting equation reveals that the integral representing the eddy buoyancy flux divergence vanishes by symmetry if the excitation is reflection symmetric. 

Substituting (\ref{eq:PsiSolution}) into (\ref{eq:Ufinal}) and simplifying we obtain the expression
\begin{equation}
\bar{s}_U=\varepsilon \iint \mbox{d}p\mbox{d}q s_U'' \mathcal{F}_U(p,q,m,N_0^2,r,\nu,s_U) \tilde{E}(p,q), \label{eq:dispersionUlong}
\end{equation}
in which we have defined $s''_{U,B}=s_{U,B}+2(r+\nu(h^2+m(q+m/2)))$. 
The feedback factor $\mathcal{F}_U$ is given by
\begin{multline}
\mathcal{F}_U = \frac{m p^2 h^2 \left(q+\frac{m}{2}\right)}{(r+\nu h^2)(2\pi)^2}\times \\
 \frac{(1-\frac{m^2}{h^2})[s_U''^2h^2h_{++}^2+2p^2N_0^2(h^2+m(q+\frac{m}{2}))]+2p^2h_{++}^2N_0^2}{[s_U''^2h^2h_{++}^2+2p^2N_0^2(h^2+m(q+\frac{m}{2}))]^2-4p^4N_0^4h^2h_{++}^2}, \label{eq:FUlong}
\end{multline}
where $h_{++}^2=p^2+(q+m)^2$. 
If $\nu=0$ the factor $s_U''$  in (\ref{eq:dispersionUlong}) may be brought to the left-hand side and we obtain the dispersion relation in the form (\ref{eq:dispersion_U}). 

To obtain the dispersion relation for the buoyancy layering instability we proceed similarly, setting $\hat{U}=0$ in (\ref{eq:Ztildefinal})-(\ref{eq:Bfinal}) so that the horizontal mean structure corresponds to buoyancy layers with no mean flow perturbation.  
Solving equations (\ref{eq:Ztildefinal})-(\ref{eq:Ktildeminusfinal}) for $\tilde{S}^{\text{diff}}$, substituting into (\ref{eq:Bfinal}), and simplifying, we obtain
\begin{equation}
\bar{s}_B=\varepsilon \int \text{d}p\text{d}q s_B'' \mathcal{F}_B (p,q,m,N_0^2,r,\nu,s_B) \tilde{E}(p,q). \label{eq:dispersionBlong}
\end{equation}
The feedback factor $\mathcal{F}_B$ is given by
\begin{equation}
\mathcal{F}_B = \frac{h^2 m^2 p^2h_{++}^4}{2(r+\nu h^2)(2\pi)^2}\frac{s_B''^2h^2h_{++}^2+2p^2m(q+m/2)N_0^2}{m^2(q+m/2)^2F_1^2-F_2F_3}, \label{eq:FBlong}
\end{equation}
in which the functions $F_{1,2,3}$ are given by
\begin{align}
F_1 &= h^2h_{++}^2s_B''^2+2p^2N_0^2(h^2+m(q+m/2)), \\
F_2 &= h^2h_{++}^2(h^2+m(q+m/2))s_B''^2+2m^2p^2(q+m/2)^2N_0^2, 
\end{align}
\begin{multline}
F_3 = h^2h_{++}^2 (h^2+m(q+\frac{m}{2}))s_B''^2 \\ +2p^2N_0^2((h^2+m(q+\frac{m}{2}))^2+h^2h_{++}^2).
\end{multline}
As in the VSHF case, the assumption that $\hat{U}=0$ can be shown to be consistent by solving (\ref{eq:Ztildefinal})-(\ref{eq:Ktildeminusfinal}) for $\tilde{\Psi}$, substituting the result into (\ref{eq:Ufinal}), and verifying that the integral on the right-hand side of the resulting equation, which represents the eddy momentum flux divergence, vanishes by symmetry for excitation satisfying reflection symmetry. 

\appendix[B] \label{sec:AppendixIRF}
\appendixtitle{Analysis of the VSHF-forming instability in the case of Isotropic Ring Excitation (IRE)}
\vspace{3mm}

\textit{(i) Mathematical formulation of IRE} 

Isotropic ring excitation is defined by the excitation spectra
\begin{align}
\tilde{\Xi}(p,q) = 2\pi k_e \delta(h-k_e), && \tilde{\Theta}(p,q) = 2\pi N_0^2 k_e^{-1} \delta(h-k_e), \label{eq:IREspectral}
\end{align}
with $\tilde{G}^{\zeta}=\tilde{G}^b=0$. 
The excitation (\ref{eq:IREspectral}) satisfies the equal-energy and non-correlation conditions (\ref{eq:ExcitationAssumption}). 
In physical space, the excitation covariances are given by
\begin{align}
\Xi(x,z) = k_e^2 J_0 (k_e r), && \Theta(x,z) = N_0^2 J_0 (k_e r),
\end{align}
where $r=\sqrt{x^2+z^2}$ and $J_0$ is the zeroth-order Bessel function of the first kind. 
Hereafter we work in nondimensional units in which the unit of length is set by the excitation scale, $1/k_e$, and the unit of time is set by the perturbation damping time, $1/r$. 
In these units the energy excitation spectrum for IRE is given by (\ref{eq:varepsIRE}). \\

\textit{(ii) Dispersion relation and feedback factor} 

To obtain explicit expressions for the dispersion relation and feedback factor for the VSHF-forming instability we evaluate equations (\ref{eq:dispersionUlong}) and (\ref{eq:FUlong}) with the excitation spectrum (\ref{eq:varepsIRE}). 
We obtain
\begin{widetext}
\begin{multline}
\bar{s}_U = \varepsilon \int_0^{2\pi}\text{d}\theta (2\pi s_U'') \left\{ \frac{m \cos^2\theta (\sin \theta + m/2)}{(2\pi)^2(1+\nu)} \right. \\
\left. \times \frac{s_U''^2 (1-m^2)(1+2m\sin\theta+m^2)+N_0^2\cos^2\theta[4+m^2(1-m^2)-2m(m^2-3)\sin\theta]}{[s_U''^2(1+2m\sin\theta+m^2)+\cos^2\theta N_0^2(2+2m\sin\theta+m^2)]^2-4\cos^4\theta N_0^4(1+2m\sin\theta+m^2)} \right\}. \label{eq:dispersionUIRElong}
\end{multline}
\end{widetext}
In this equation the explicit nondimensional forms of $\bar{s}_U$ and $s_U''$ for IRE are given by $\bar{s}_U=s_U+r_m+\nu m^2$ and $s_U''=s_U+2(1+\nu(1+m(\sin\theta+m/2)))$. 
The factor in braces in (\ref{eq:dispersionUIRElong}) gives the explicit form of the feedback factor, $\mathcal{F}_U$, that is relevant to the case of IRE as discussed in Section \ref{sec:feedback}. 
Note that the factor of $2\pi$ outside the braces originates from (\ref{eq:varepsIRE}) and so is not included in $\mathcal{F}_U$. 
Explicit formulae for the buoyancy layering dispersion relation and feedback factor can be obtained by a similar procedure in which equations (\ref{eq:dispersionBlong}) and (\ref{eq:FBlong}) are evaluated using the excitation spectrum (\ref{eq:varepsIRE}). \\

\textit{(iii) Asymptotic analysis}


We now provide details of the derivations of various asymptotic approximations useful for understanding the properties of the VSHF-forming instability in the case of IRE. 
For simplicity we set $\nu=0$ throughout. 

To obtain the estimate (\ref{eq:IREsmallNsqsU}) for the VSHF growth rate under weak stratification, we first note that in the case $N_0^2=0$ the dispersion relation (\ref{eq:dispersionUIRElong}) simplifies to
\begin{equation}
\bar{s}_U s'_U = \varepsilon m (1-m^2) \int_0^{2\pi} \frac{\text{d}\theta}{2\pi}\frac{\cos^2\theta(\sin\theta+m/2)}{1+2m\sin\theta+m^2}. \label{eq:IREintegral1}
\end{equation}
The integral on the right-hand side of (\ref{eq:IREintegral1}) is equal to zero for $0<m<1$, and the VSHF growth rate is then given by $s_U=-r_m$.
To obtain the leading order correction to this $N_0^2=0$ solution we substitute the expansion $s_U = -r_m + s_1 N_0^2 + \mathcal{O}(N_0^4)$ into (\ref{eq:dispersionUIRElong}) and expand the integrand in a power series in $N_0^2$, retaining terms up to order $N_0^2$. 
This procedure gives a number of integrals similar in form to the integral in (\ref{eq:IREintegral1}), all of which can be evaluated in closed form. 
Solving the resulting expression for $s_1$ gives the result (\ref{eq:IREsmallNsqsU}).

To obtain the expressions (\ref{eq:eddy_viscosity_1})-(\ref{eq:eddy_viscosity_2}), which illustrate that the VSHF-forming instability is associated with negative eddy viscosity, we analyze the dispersion relation (\ref{eq:dispersionUIRElong}) in the limit of very large-scale VSHFs, corresponding to small $m$. 
For $m=0$ the VSHF growth rate is given by the explicit damping rate, $s_U=-r_m$, which can be verified by inspection of (\ref{eq:dispersionUIRElong}). 
For small $m$ we write the instability growth rate as $s_U = -r_m + s_1 m^2+\mathcal{O}(m^4)$, omitting terms of odd order as $s_U$ does not depend on the sign of $m$.
Substituting this expression into (\ref{eq:dispersionUIRElong}) and retaining terms in the expansion up to order $m^2$ gives a sum of integrals that can be evaluated in closed form. 
Solving the resulting expression for $s_1$ gives
\begin{equation}
s_1 = \varepsilon g(N_0^2,r_m), \label{eq:eddy_viscosity_s1}
\end{equation}
where $g$ is defined by 
\begin{equation}
g(N_0^2,r_m) = \frac{1}{16s_0'}\left\{1-\frac{s_0'^2}{N_0^2}(1-2f)-\frac{s_0'^4}{2N_0^4}(1-f) \right\},
\end{equation}
in which we have used the notation $s_0'=2-r_m$ and defined the function $f(r_m,N_0^2)=s_0'(4N_0^2+s_0'^2)^{-1/2}$. 
Equation (\ref{eq:eddy_viscosity_1}) identifies the eddy viscosity, $\nu_{eddy}$, with the negative of the growth rate correction, $-s_1$, which gives the result (\ref{eq:eddy_viscosity_2}). 
Analysis of $g(N_0^2,r_m)$ reveals that $g>0$ for all $N_0^2$, so that the eddy viscosity is negative.
%
%

To obtain the estimate (\ref{eq:IRE_LargeNsq_sU}) of the VSHF growth rate in the case of strong stratification we analyze the dispersion relation (\ref{eq:dispersionUIRElong}) in the limit of large $N_0^2$. 
As $N_0^2\to\infty$, inspection of (\ref{eq:dispersionUIRElong}) shows that $s_U\to-r_m$. 
To obtain the leading-order correction for large but finite $N_0^2$, we write the dispersion relation terms of the small parameter $\delta \equiv 1/N_0^2$ as
\begin{equation}
\bar{s}_U= m\varepsilon \delta s_0' \int \frac{\text{d}\theta}{2\pi} I(\delta,\theta), \label{eq:IRElargeNsq1}
\end{equation}
where the integrand is given by 
\begin{widetext}
\begin{equation}
I = \frac{ \cos^4\theta (\sin\theta+m/2) [4+m^2(1-m^2)-2m(m^2-3)\sin\theta]+\mathcal{O}(\delta)}{m^2 \cos^4\theta (2\sin\theta+m)^2+\delta [2s_0'^2\cos^2\theta(1+2m\sin\theta+m^2)(2+2m\sin\theta+m^2)]+\mathcal{O}(\delta^2)}.
\end{equation}
\end{widetext}
\hspace{-2mm}The factor of $\delta$ outside the integral in (\ref{eq:IRElargeNsq1}) indicates that the correction to the growth rate decays at least as fast as $1/N_0^2$.
To obtain the explicit form of the correction, we evaluate the integral using the residue theorem. 
The integral of $I(\delta,\theta)$ is undefined for $\delta=0$ due to the presence of poles at the solutions of $2\sin\theta+m=0$ which exist when $0<m<2$. 
These poles are shifted into the complex plane for small but nonzero $\delta$ by the $\mathcal{O}(\delta)$ term in the denominator. 
To apply the residue theorem we use a rectangular contour in the complex plane that includes the real interval $[0,2\pi]$ and is closed in the upper half plane. 
To use this contour we must ensure that the integrand, $I(\delta,\theta)$, vanishes as $\theta\to +i\infty$. 
However, $I$ does not vanish in this limit and instead converges to 
\begin{equation}
\lim_{\theta \to +i\infty} I(\delta,\theta) = \frac{3-m^2}{2m},
\end{equation}
and so the residue theorem cannot be applied directly to (\ref{eq:IRElargeNsq1}). 
This issue is resolved by adding and subtracting this limiting value inside the integral in (\ref{eq:IRElargeNsq1}) to obtain
\begin{equation}
\bar{s}_U = m\varepsilon \delta s_0' \int \frac{d\theta}{2\pi} \left[ I(\delta,\theta) - \frac{3-m^2}{2m} \right] + \frac12 \varepsilon \delta s_0' (3-m^2). \label{eq:IRElargeNsq2}
\end{equation}
The integral in (\ref{eq:IRElargeNsq2}) can be evaluated using the residue theorem. 
Due to the $2\pi$-periodicity of the integrand, the vertical branches of the contour integral cancel one another, and detailed calculation of the residues shows that the contributions from the poles also sum to zero so that the integral in (\ref{eq:IRElargeNsq2}) equals zero at the lowest order in $\delta$. 
The growth rate estimate (\ref{eq:IRE_LargeNsq_sU}) is then obtained by solving (\ref{eq:IRElargeNsq2}) for $s_U$. 

To obtain asymptotic estimates for the stability boundary, $\varepsilon_c(m)$, in the limits of weak and strong stratification, we follow identical procedures for expanding and evaluating the integral in (\ref{eq:stabilityboundary}) as were used to obtain the asymptotic growth rate estimates in those limits, except that we set $s_U=0$ rather than expanding about $s_U=-r_m$. 
In the limit of weak stratification we obtain 
\begin{equation}
\varepsilon_c(m) \approx \frac{64 r_m}{m^2 N_0^2}, \label{eq:IREstabbnd1}
\end{equation}
which is valid for $m<1$. 
The first VSHF wavenumber to become unstable is then $m^{\star}=1$, and so the stability boundary is given by $\varepsilon_c \approx 64 r_m / N_0^2$. 
In the limit of strong stratification we obtain
\begin{equation}
\varepsilon_c(m) \approx \frac{r_m N_0^2}{3-m^2}
\end{equation}
which is valid for $m<\sqrt{3}$, as $\varepsilon$ must be positive. 
In this case the first VSHF wavenumber to become unstable tends to $m^{\star}=0$ as $N_0^2\to \infty$, and so the stability boundary is given by $\varepsilon_c \approx r_m N_0^2/3$. 
These estimates are shown in Figure \ref{fig:StabilityBoundaryAsymptotics} (a).

\appendix[C] \label{sec:AppendixMCE}
\appendixtitle{Analysis of the VSHF-forming instability in the case of Monochromatic Excitation (MCE)}

\textit{(i) Dispersion relation}

MCE is defined by the energy injection spectrum (\ref{eq:varepsMCE}), which corresponds to the vorticity and buoyancy excitation spectra 
\begin{align}
\tilde{\Xi}&= \pi^{3/2}\ell_c (1+q^2) \exp (-\ell_c^2 q^2/4) \left[\delta(p+1)+\delta(p-1) \right] \\
\tilde{\Theta} & = N_0^2 \pi^{3/2}\ell_c \exp (-\ell_c^2 q^2/4) \left[\delta(p+1)+\delta(p-1) \right]
\end{align}
As in the case of IRE, we obtain the dispersion relation for the VSHF-forming instability in the case of MCE by evaluating (\ref{eq:dispersionUlong}) and (\ref{eq:FUlong}) with the excitation spectrum (\ref{eq:varepsMCE}). 
We obtain
\begin{widetext}
\begin{multline}
\bar{s}_U = \varepsilon \int_{-\infty}^{\infty} \text{d}q s_U'' \left( 2\pi^{3/2}\ell_c e^{-\ell_c^2 q^2/4}\right) \Bigg\{ \frac{m(q+m/2)(1+q^2)}{(2\pi)^2(1+\nu(1+q^2))} \times \\
\frac{\left(1-\frac{m^2}{1+q^2}\right)\left[s_U''^2(1+q^2)(1+(q+m)^2)+2N_0^2(1+q^2+m(q+m/2)) \right] + 2N_0^2 (1+(q+m)^2)}{\left[ s_U''^2(1+q^2)(1+(q+m)^2)+2N_0^2(1+q^2+m(q+m/2)) \right]^2-4N_0^4(1+q^2)(1+(q+m)^2)}\Bigg\}, \label{eq:dispersionUMCElong}
\end{multline}
\end{widetext}
\hspace{-2mm}in which $\bar{s}_U=s_U+r_m+\nu m^2$ and $s_U''=s_U + 2(1+\nu(1+q^2+m(q+m/2)))$. 
The factor in braces in (\ref{eq:dispersionUMCElong}) gives the explicit form of the feedback factor, $\mathcal{F}_U$, that is relevant to the case of MCE. 
Explicit formulae for the buoyancy layering dispersion relation and feedback factor in the case of MCE can be obtained by evaluating (\ref{eq:dispersionBlong}) and (\ref{eq:FBlong}) using the excitation spectrum (\ref{eq:varepsMCE}). \\

\textit{(ii) Asymptotic analysis}

To obtain a closed form estimate of the growth rate of the VSHF-forming instability in the case of MCE under strong stratification, we expand (\ref{eq:dispersionUMCElong}) in the small parameter $\delta=1/N_0^2$ to obtain
\begin{equation}
\bar{s}_U=\delta \frac{\varepsilon \ell_c s'_0}{2m\sqrt{\pi}}\int_{-\infty}^{\infty}\text{d}q e^{-\ell_c^2 q^2/4}\frac{J(q)}{(m+2q)+\mathcal{O}(\delta)}. \label{eq:dispersionUMCElargeNsq1}
\end{equation}
where we have set $\nu=0$ for simplicity and defined $s'_0=2-r_m$ and $J(q) = (1+(q+m)^2)(1+q^2)+(1+q^2+m(q+m/2))(1+q^2-m^2)$. 
As in the case of IRE, the factor of $\delta$ outside the integral indicates that $\bar{s}_U$ decays at least as fast as $1/N_0^2$ as the stratification is increased. 
The integral in (\ref{eq:dispersionUMCElargeNsq1}) is undefined for $\delta=0$ due to the pole at $q=-m/2$. 
However, the value of the integral converges to a well-defined limit as $\delta\to0$, which can be evaluated as follows. 
To regularize the $\delta=0$ integral at $q=-m/2$, we rewrite (\ref{eq:dispersionUMCElargeNsq1}) as 
\begin{multline}
\bar{s}_U=\delta \frac{\varepsilon \ell_c s'_0}{2m\sqrt{\pi}} \Bigg\{ \int_{-\infty}^{\infty}\text{d}q e^{-\ell_c^2 q^2/4}\frac{J(q)-J(-m/2)}{(m+2q)+\mathcal{O}(\delta)} \\
+J(-m/2) \int_{-\infty}^{\infty}\text{d}q \frac{e^{-\ell_c^2 q^2/4}}{(m+2q)+\mathcal{O}(\delta)} \Bigg\}\label{eq:dispersionUMCElargeNsq2}
\end{multline}
The first integral in (\ref{eq:dispersionUMCElargeNsq2}) is no longer singular at $q=-m/2$ for $\delta=0$ and can be evaluated in closed form. 
The second integral remains singular at $\delta=0$. 
However, it can be assigned a finite Cauchy principal value as $\delta\to0$, and in fact can be recognized, after minor manipulations, as the Hilbert transform of a Gaussian function. 
We then obtain
\begin{equation}
s_U \approx -r_m +(1-r_m/2)\frac{\ell_c \varepsilon}{N_0^2} S(m),
\end{equation}
where the function $S$ is given by 
\begin{equation}
S(m) = \frac{4+\ell_c^2}{\ell_c^3}-\frac{3}{4\ell_c}m^2+\left(\frac{2}{m}-\frac{m^3}{8}\right)F\left(\frac{\ell_c m}{4}\right),
\end{equation}
in which $F$ is the Dawson function. 
This approximation is shown in Figure \ref{fig:LargeNsqAsymp}. 

An asymptotic estimate for the stability boundary, $\varepsilon_c$, can be obtained by applying similar methods to approximate the integral in (\ref{eq:stabilityboundary}). 
In the limit of strong stratification we obtain
\begin{equation}
\varepsilon_c(m) \approx \frac{r_m N_0^2}{\ell_c S(m)}.
\end{equation}
In this limit the minimum of $\varepsilon_c(m)$ occurs at $m^{\star}=0$, so that the stability boundary is given by \begin{equation}
\varepsilon_c \approx \frac{ 2 r_m \ell_c^2}{\ell_c^4+2\ell_c^2+8}N_0^2.
\end{equation}
This estimate is shown in Figure \ref{fig:StabilityBoundaryAsymptotics} (a).

\appendix[D] \label{sec:AppendixMechanism}
\appendixtitle{Decomposition of the Feedback Factor into Contributions from Individual Processes}

In this Appendix we provide mathematical details relevant to Section \ref{sec:processes} in which the feedback factor for the VSHF-forming instability, $\mathcal{F}_U$, is decomposed into the feedback contributions from individual processes. 
For simplicity we set $\nu=0$. 

When solving (\ref{eq:Ztildefinal})-(\ref{eq:Ktildeminusfinal}) for $\tilde{\Psi}$ as described in Appendix A, the solution can be decomposed as 
\begin{equation}
\tilde{\Psi} = \tilde{\Psi}^{Orr} + \tilde{\Psi}^{cu} + \tilde{\Psi}^{wave}, \label{eq:MechDecomp00}
\end{equation}
where the individual contributions are defined by the term in (\ref{eq:S3T_lin_phys_1_MAIN}) from which they each originate, as described in Section \ref{sec:processes}. 
The contributions are given by
\begin{align}
\tilde{\Psi}^{Orr} &= -ip\hat{U} \frac{h_{-}^4\tilde{\Psi}_H^{-}-h_{+}^4\tilde{\Psi}_H^{+}}{s'_Uh_{-}^2h_{+}^2}, \\
\tilde{\Psi}^{cu} &= im^2 p \hat{U} \frac{h_{-}^2\tilde{\Psi}_H^{-}-h_{+}^2\tilde{\Psi}_H^{+}}{s'_Uh_{-}^2h_{+}^2}.
\end{align}
The contribution from wave dynamics, $\tilde{\Psi}^{wave}$, can be obtained most simply as a residual using (\ref{eq:MechDecomp00}) and (\ref{eq:PsiSolution}). 
Combining the decomposition (\ref{eq:MechDecomp00}) with equations (\ref{eq:Ufinal}) and (\ref{eq:dispersionUlong}) yields the decomposition (\ref{eq:FU_decomp}). 


\bibliographystyle{ametsoc2014}

\bibliography{Analytical_Bibfile_Refined}

\end{document}